\documentclass[runningheads]{llncs}

\usepackage{graphicx}
\usepackage{graphics}
\usepackage{wrapfig}
\usepackage{float}
\usepackage{textcomp}

\usepackage{graphicx}
\begin{document}
\title{Model-based Learning for Quantitative Susceptibility Mapping}
\titlerunning{Model-based Learning for QSM}
%
\author{Juan Liu\inst{1}\orcidID{0000-1111-2222-3333} \and
Kevin M. Koch\inst{2,3}\orcidID{1111-2222-3333-4444}}
\authorrunning{J. Liu et al.}
%

\author{{Juan Liu\inst{1,2}} \and
Kevin M. Koch\inst{1,2,3}}
\authorrunning{J. Liu, K.M. Koch}
%
\institute{Center for Imaging Research, Medical College of Wisconsin, Milwaukee, WI, USA \and
Department of Biomedical Engineering, Marquette University and Medical College of Wisconsin, Milwaukee, WI, USA \and
Department of Radiology, Medical College of Wisconsin, Milwaukee, WI, USA\\
\email{juan.liu@marquette.edu}}

\maketitle              
\begin{abstract} 
Quantitative susceptibility mapping (QSM) is a magnetic resonance imaging (MRI) technique that estimates magnetic susceptibility of tissue from Larmor frequency offset measurements. The generation of QSM requires solving a challenging ill-posed field-to-source inversion problem. Inaccurate field-to-source inversion often causes large susceptibility estimation errors that appear as streaking artifacts in the QSM, especially in massive hemorrhagic regions. Recently, several deep learning (DL) QSM techniques have been proposed and demonstrated impressive performance. Due to the inherent non-existent ground-truth QSM references, these DL techniques used either calculation of susceptibility through multiple orientation sampling (COSMOS) maps or synthetic data for network training. Therefore, they were constrained by the availability and accuracy of COSMOS maps, or suffered from performance drop when the training and testing domains were different. To address these limitations, we present a model-based DL method, denoted as uQSM. Without accessing to QSM labels, uQSM is trained using the well-established physical model. When evaluating on multi-orientation QSM datasets, uQSM achieves higher levels of quantitative accuracy compared to TKD, TV-FANSI, MEDI, and DIP approaches. When qualitatively evaluated on single-orientation datasets, uQSM outperforms other methods and reconstructed high quality QSM.  

\end{abstract}

\section{Introduction}
Quantitative susceptibility mapping (QSM) can estimate tissue magnetic susceptibility values from magnetic resonance imaging (MRI) Larmor frequency sensitive phase images \cite{wang2015quantitative}. Biological tissue magnetism can provide useful diagnostic image contrast and be used to quantify biomarkers including iron, calcium, and gadolinium \cite{wang2015quantitative}. To date, all QSM methods rely on a dipolar convolution that relates susceptibility sources to induced Larmor frequency offsets \cite{salomir2003fast,marques2005application}, which is expressed in the k-space as bellow. 

\begin{equation} 
B(\vec k)  = X (\vec k) \cdot {D}(\vec k); {D}(\vec k) =  \frac{1}{3} - \frac{{k_z^2}}{{{k_x^2 + k_y^2 + k_z^2}}}
\end{equation}

where $B(\vec k)$ is the susceptibility induced magnetic perturbation along the main magnetic field direction, $X(\vec k)$ is the susceptibility distribution $\chi$ in the k space, $D(\vec k)$ is the dipole kernel. While the forward relationship of this model (source to field) is well-established and can be efficiently computed using Fast-Fourier-Transform (FFT), the k-space singularity in the dipole kernel results in an ill-conditioned relationship in the field-to-source inversion.

Calculation of susceptibility through multiple orientation sampling (COSMOS) \cite{liu2009calculation} remains the empirical gold-standard of QSM, as the additional field data sufficiently improves the conditioning of the inversion algorithm. Since it is time-consuming and clinically infeasible to acquire multi-orientation data, single-orientation QSM is preferred which is computed by either thresholding of the convolution operator \cite{shmueli2009magnetic,wharton2010susceptibility,haacke2010susceptibility} or use of more sophisticated regularization methods \cite{de2008quantitative,de2010quantitative,liu2011morphology,bilgic2014fast}. In single-orientation QSM, inaccurate field-to-source inversion often causes large susceptibility estimation errors that appear as streaking artifacts in the QSM, especially in massive hemorrhagic regions. 

Recently, several deep learning (DL) approaches have been proposed to solve for the QSM dipole inversion. QSMnet \cite{yoon2018quantitative} used COSMOS results as QSM labels for training, which reconstructed COSMOS-like QSM estimates no matter the head orientations. DeepQSM \cite{bollmann2019deepqsm} used synthetic susceptibility maps simulated using basic 3D geometric shapes and the forward dipole model to generate synthetic training data. QSMGAN \cite{chen2019qsmgan} adopted COSMOS maps as QSM labels and refined the network using the Wasserstein Generative Adversarial Networks (WGAN) \cite{goodfellow2014generative,arjovsky2017wasserstein}. QSMnet+ \cite{jung2020exploring} employed data augmentation approaches to increase the range of susceptibility, which improved the linearity of susceptibility measurement in clinical situations. 

Though these DL techniques have exhibited impressive results, there were several limitations. These methods are supervised and data-driven which require QSM labels for network training. Unfortunately, QSM has the inherent non-existent `ground-truth'. Therefore, these methods used either COSMOS data or synthetic data for network training. However, acquiring a large number of COSMOS data is not only expensive but also time consuming. Moreover, COSMOS neglects tissue susceptibility anisotropy \cite{liu2010susceptibility} and contains errors from background field removal and image registration procedures, which compromises COSMOS map as a QSM label. Though synthetic data provides a reliable and cost-effective way for training, the generalization capability needs to be addressed since the domain gap between the synthetic training data and real data often causes performance degradation and susceptibility quantification errors.

Here, we propose a model-based learning method without the need of QSM labels for QSM dipole inversion, denoted as uQSM, to overcome these limitations. Quantitative evaluation is performed on multi-orientation datasets in comparison to TKD \cite{shmueli2009magnetic}, TV-FANSI \cite{milovic2018fast}, MEDI \cite{liu2012morphology}, and deep image prior (DIP)\cite{ulyanov2018deep}, with COSMOS result as a reference. In addition, qualitative evaluation is performed on single-orientation datasets.

\begin{figure}[H]
\vspace{-15pt}
\begin{center}
\includegraphics[width=\textwidth]{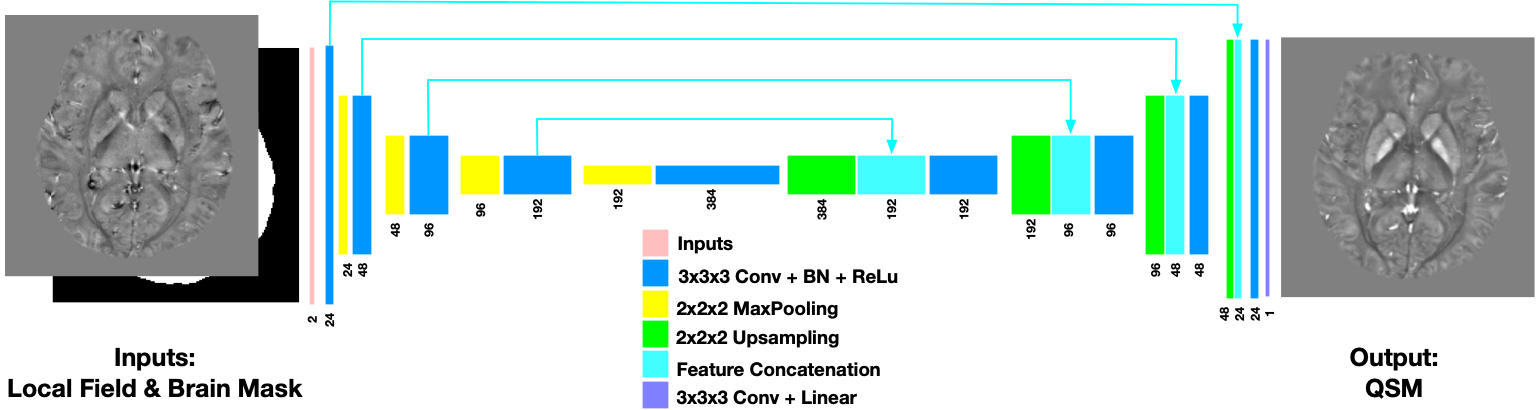}
\caption{Neural network architecture of uQSM. It has an encoder-decoder structure with 9 convolutional layers (kernel size 3x3x3, same padding), 9 batch normalization layers, 9 ReLU layers, 4 max pooling layers (pooling size 2x2x2, strides 2x2x2), 4 nearest-neighbor upsampling layers (size 2x2x2), 4 feature concatenations, and 1 convolutional layer (kernel size 3x3x3, linear activation).}
\label{fig:uQSM_cnn}
\vspace{-20pt}
\end{center}
\end{figure}

\section{Method}

uQSM adopted a 3D convolutional neural network with an encoder-decoder architecture as shown in Fig.\ref{fig:uQSM_cnn}. 

The network took two inputs, a local field measurement $f$ and a brain mask $m$, and one output, a susceptibility map $\chi$. Upsampling layers were used in the decoding path in stead of deconvolutional layers to address the checkboard artifacts \cite{odena2016deconvolution} in the reconstructed QSM. 

The loss function incorporated the model-based data consistency loss $L_{\chi}$. 

\begin{equation}
L_{\chi} = \left \| W  m(e^{j d\ast \chi}  - e^{j f}) \right \|_{2}
\end{equation}

where $W$ serves as a data-weighting factor which can be the magnitude image or noise weight matrix, $d$ is the dipole kernel, $\ast$ is the convolution operator. Since noise is unknown and spatially variant in the local field measurements, the nonlinear dipole convolution data consistency loss was used in the loss function to get more robust QSM estimates as conventional QSM methods \cite{liu2013nonlinear,polak2020nonlinear}. The dipole convolution was computed in the k-space using FFT. 

For normalization, the mean and standard deviation were calculated in the local fields. Then, the input local maps were normalized to have a mean of 0 and a standard deviation of 1. Since the susceptibility maps were unknown, we used $3 d\ast \chi$ to be consistent with $f$, which can make the susceptibility outputs close to being standard normalized. 

\begin{equation}
L_{TV} = \left \| G_{x}(\chi )\right \|_{1} + \left \| G_{y}(\chi )\right \|_{1} + \left \| G_{z}(\chi )\right \|_{1}
\end{equation}

In addition, a total variation (TV) loss $L_{TV}$ was included to serve as a regularization term to preserve important details such as edges whilst removing unwanted noise in the reconstructed susceptibility maps. In $L_{TV}$, $G_{x}$, $G_{y}$, $G_{z}$ are gradient operators in x, y, z directions. 

\begin{equation}
L_{Total} = L_{\chi} + \lambda L_{TV}
\label{uQSM_loss}
\end{equation}

The loss function is the weighted sum of the data consistency loss $L_{\chi}$ and the total variation loss $L_{TV}$. 


\section{Experiments}

\textbf{Multi-orientation QSM Data} 
9 QSM datasets were acquired using 5 head orientations and a 3D single-echo GRE scan with isotropic voxel size 1.0x1.0x1.0 mm$^3$ on 3T MRI scanners. QSM data processing was implemented as following, offline GRAPPA \cite{griswold2002generalized} reconstruction to get magnitude and phase images from saved k-space data, coil combination using sensitivities estimated with ESPIRiT \cite{uecker2014espirit}, BET (FSL, FMRIB, Oxford, UK) \cite{smith2002fast} for brain extraction, Laplacian method \cite{li2011quantitative} for phase unwrapping, and RESHARP \cite{wu2012whole} with spherical mean radius 4mm for background field removal. COSMOS results were calculated using the 5 head orientation data which were registered by FLIRT (FSL, FMRIB, Oxford, UK) \cite{jenkinson2002improved,jenkinson2001global}. In addition, QSM estimates at the normal head position were generated using the TKD, TV-FANSI, and MEDI algorithms. 

For uQSM, leave-one-out cross validation was used. For each dataset, total 40 scans from other 8 datasets were used for training. uQSM was trained using patch-based neural network with patch size 96x96x96. The RESHARP local field and brain mask patches with patch size 96x96x96 were cropped with an overlapping scheme of 16.6 percent overlap between adjacent patches. These patch pairs were used for training and validation with split ratio 9:1. The magnitude images were scaled between 0 to 1 and used as the weighting factor $W$, $\lambda$ was set 0.001. The Adam optimizer \cite{kingma2014adam} was used for the model training. The initial learning rate was set as 0.0001, with exponentially decay at every 100 steps. One NVIDIA GPU Tesla k40 was used for training with batch size 4. The model was trained and evaluated using Keras with Tensorflow as a backend. After training, the full local field and brain mask from the leave-one dataset were fed to the trained model to get the QSM estimates.  

In addition, we used DIP to get the QSM images. DIP used the same neural network architecture and loss function described above. DIP was performed on each individual dataset using full neural network. To avoid overfitting which can introduce artifacts in the reconstructed QSM images, DIP was stopped after 200 iterations to get QSM results. 

The QSM of uQSM, TKD, TV-FANSI, MEDI, and DIP were compared with respect to the COSMOS maps using quantitative metrics, peak signal-to-noise ratio (pSNR), normalized root mean squared error (NRMSE), high frequency error norm (HFEN), and structure similarity (SSIM) index. 

\textbf{Single-orientation QSM Data} 
150 QSM datasets were collected on a 3T MRI scanner (GE Healthcare MR750) from a commercially available susceptibility weighted software application (SWAN, GE Healthcare). The data acquisition parameters were as follows: in-plane data matrix 320x256, field of view 24 cm, voxel size 0.5x0.5x2.0 mm$^3$, 4 echo times [10.4, 17.4, 24.4, 31.4] ms, repetition time 58.6 ms, autocalibrated parallel imaging factors 3x1, and total acquisition time 4 minutes.

Complex multi-echo images were reconstructed from raw k-space data using customized code. The brain masks were obtained using the SPM tool \cite{brett2002region}. After background field removal using the RESHARP with spherical mean radius 4mm, susceptibility inversion was performed using TKD, TV-FANSI, and MEDI. In addition, we used DIP to get the QSM images. DIP was performed for each individual dataset and early stopped after 200 iterations.  

For uQSM training, 10449 patch pairs of local field maps and brain masks with patch size 128x128x64 were extracted from 100 QSM datasets. The average of multi-echo magnitude images was scaled between 0 to 1 and used as the weighting factor $W$. In the loss function, $\lambda$ was set 0.0 since the data had high signal-to-noise ratio. The Adam optimizer was used with an initial learning rate 0.0001, which exponentially decayed at every 100 steps. One NVIDIA GPU Tesla k40 was used for training with batch size 2. After training, the trained DL model took full local fields and brain masks to get the QSM estimates. 

\section{ Experimental Results}

\begin{table}[H]
\vspace{-20pt}
\centering
\caption{\label{tab:uQSM1} Means and standard deviations of quantitative performance metrics of 5 reconstructed QSM images with COSMOS as a reference on 9 multi-orientation datasets.}
\vspace{0in}
\begin{tabular}{cccccc}
\hline
\multicolumn{1}{|c}{} & \multicolumn{1}{|c}{pSNR (dB)} & \multicolumn{1}{|c}{NRMSE ($\%$)} &\multicolumn{1}{|c}{HFEN ($\%$)} &\multicolumn{1}{|c|}{SSIM (0-1)}\\
\hline 
\multicolumn{1}{|c}{TKD} & \multicolumn{1}{|c}{$43.4\pm0.5$} & \multicolumn{1}{|c}{$91.4\pm6.7$} & \multicolumn{1}{|c}{$72.9\pm6.6$} & \multicolumn{1}{|c|}{$0.831\pm0.016$}\\
\hline
\multicolumn{1}{|c}{TV-FANSI} & \multicolumn{1}{|c}{$41.5\pm0.6$} & \multicolumn{1}{|c}{$80.0\pm5.0$} & \multicolumn{1}{|c}{$73.6\pm6.2$} & \multicolumn{1}{|c|}{$0.869\pm0.019$}\\
\hline
\multicolumn{1}{|c}{MEDI} & \multicolumn{1}{|c}{$41.5\pm0.6$} & \multicolumn{1}{|c}{$113.8\pm7.6$} & \multicolumn{1}{|c}{$100.4\pm9.1$} & \multicolumn{1}{|c|}{\textbf{0.902$\pm$0.016}}\\
\hline
\multicolumn{1}{|c}{DIP} & \multicolumn{1}{|c}{$44.0\pm0.8$} & \multicolumn{1}{|c}{$85.5\pm6.7$} & \multicolumn{1}{|c}{$65.7\pm4.5$} & \multicolumn{1}{|c|}{{$0.859\pm0.020$}}\\
\hline
\multicolumn{1}{|c}{uQSM} & \multicolumn{1}{|c}{\textbf{45.6$\pm$0.4}} & \multicolumn{1}{|c}{\textbf{71.4$\pm$5.0}} & \multicolumn{1}{|c}{\textbf{62.8$\pm$5.0}} & \multicolumn{1}{|c|}{$0.890\pm0.015$}\\
\hline 
\end{tabular}
\begin{flushleft}
\end{flushleft}
\vspace{-25pt}
\end{table}

\begin{figure}[H]
\begin{center}
\vspace{-20pt}
\includegraphics[width=\textwidth]{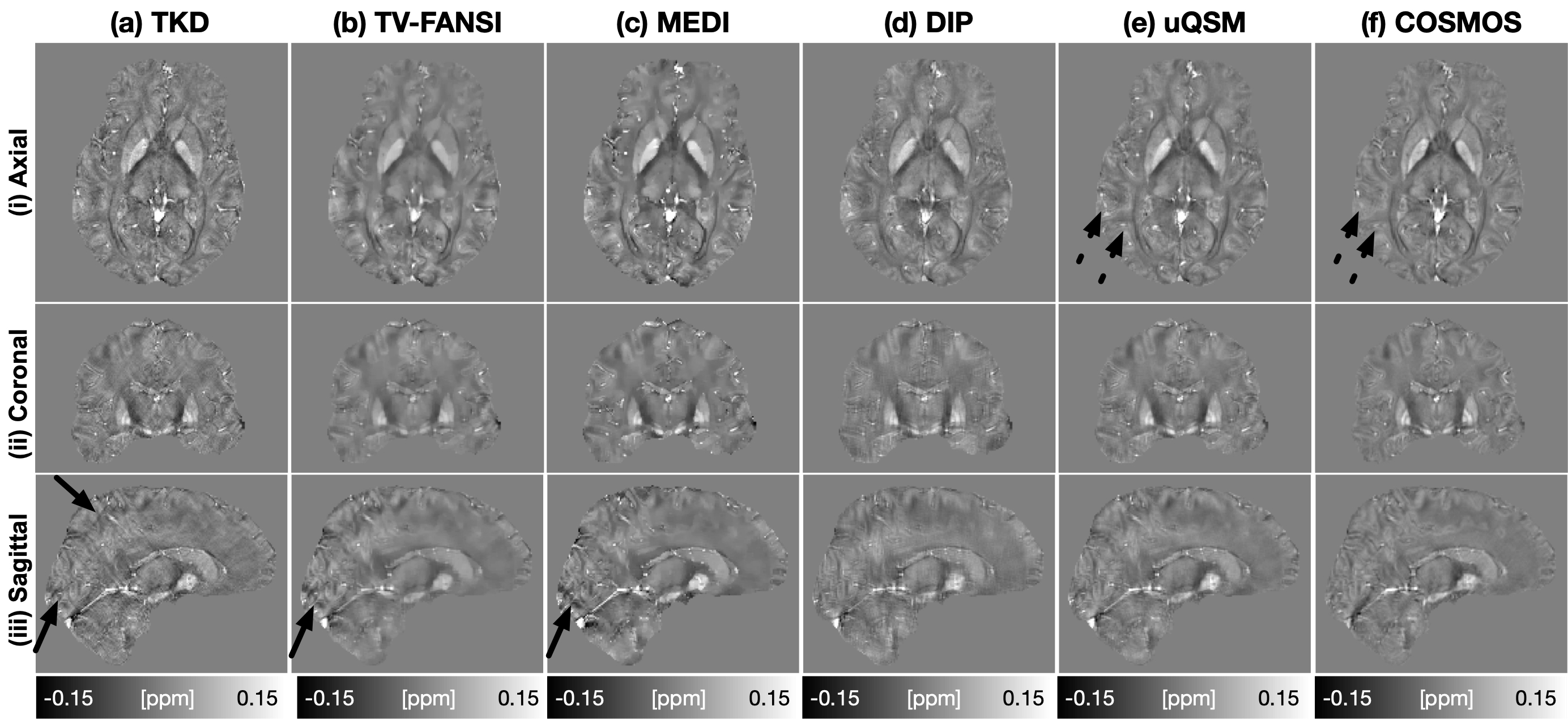}
\vspace{-15pt}
\caption{Comparison of QSM of a multi-orientation data. TKD (a), TV-FANSI (b) and MEDI (c) maps showed oversmoothing and/or streaking artifacts. The uQSM (e) maps well preserved image details and showed invisible artifacts.}
\label{fig:uQSM_QSMdata7}
\vspace{-20pt}
\end{center}
\end{figure}

\textbf{Multi-orientation QSM Data} 
Table.\ref{tab:uQSM1} summarized quantitative metrics of 5 reconstruction methods on 9 multi-orientation datasets with COSMOS map as a reference. Compared to TKD, TV-FANSI, MEDI, and DIP, uQSM results achieved the best metric scores in pSNR, RMSE, and HFEN, and second in SSIM. Fig.\ref{fig:uQSM_QSMdata7} compared QSM images from a representative dataset in three planes. Streaking artifacts were observed in the sagittal planes of TKD, TV-FANSI, and MEDI results (a-c, iii, black solid arrows). TV-FANSI and MEDI maps showed substantial blurring due to their use of spatial regularization. DIP results displayed good image quality. uQSM demonstrated superior image sharpness and invisible image artifacts. Compared with uQSM, COSMOS results displayed conspicuity loss due to image registration errors (e-f, i, black dash arrows). 

\begin{figure}[H]
\begin{center}
\vspace{-25pt}
\includegraphics[width=0.95\textwidth]{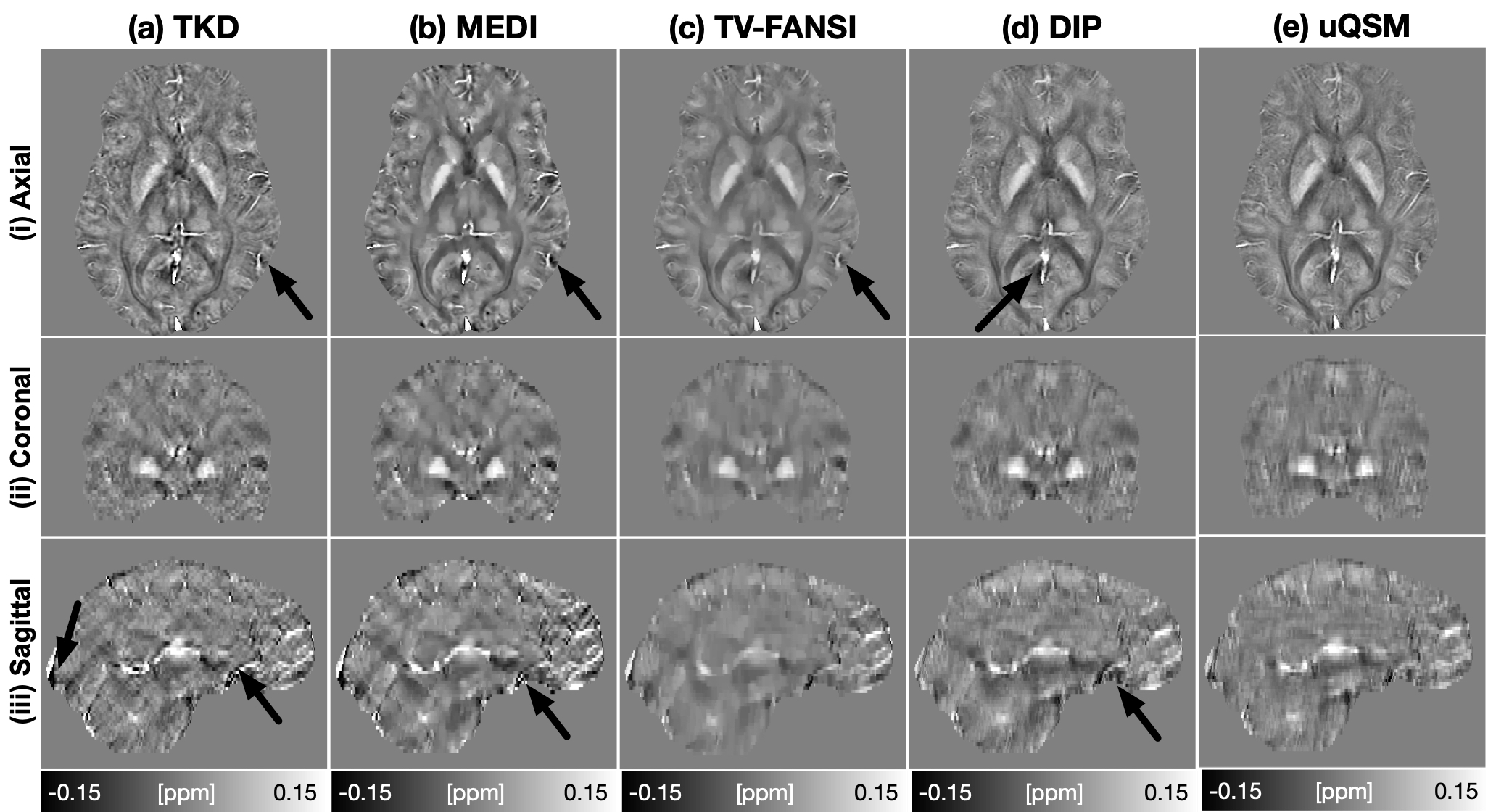}
\vspace{-5pt}
\caption{Comparison of QSM of a single-orientation dataset. TDK (a), MEDI (b), and DIP (d) results showed black shading artifacts in the axial plane and streaking artifacts in the sagittal plane. MEDI (b) and TV-FANSI (c) results suffered from oversmoothing. uQSM (e) images had high-quality  with clear details and invisible artifacts.}
\label{fig_h2h}
\vspace{-20pt}
\end{center}
\end{figure}

\textbf{Single-orientation QSM Data} 
Fig.\ref{fig_h2h} displayed QSM images from a single-orientation dataset. TKD, MEDI results had black shading artifacts in the axial plane, and streaking artifacts in coronal and sagittal planes. MEDI and TV-FANSI images showed oversmoothing and lost image sharpness. DIP results showed high quality but subtle image artifacts. Visual comparison demonstrated that uQSM outperformed other methods and produced better QSM images. 

\begin{figure}[H]
\begin{center}
\vspace{-20pt}
\includegraphics[width=0.75\textwidth]{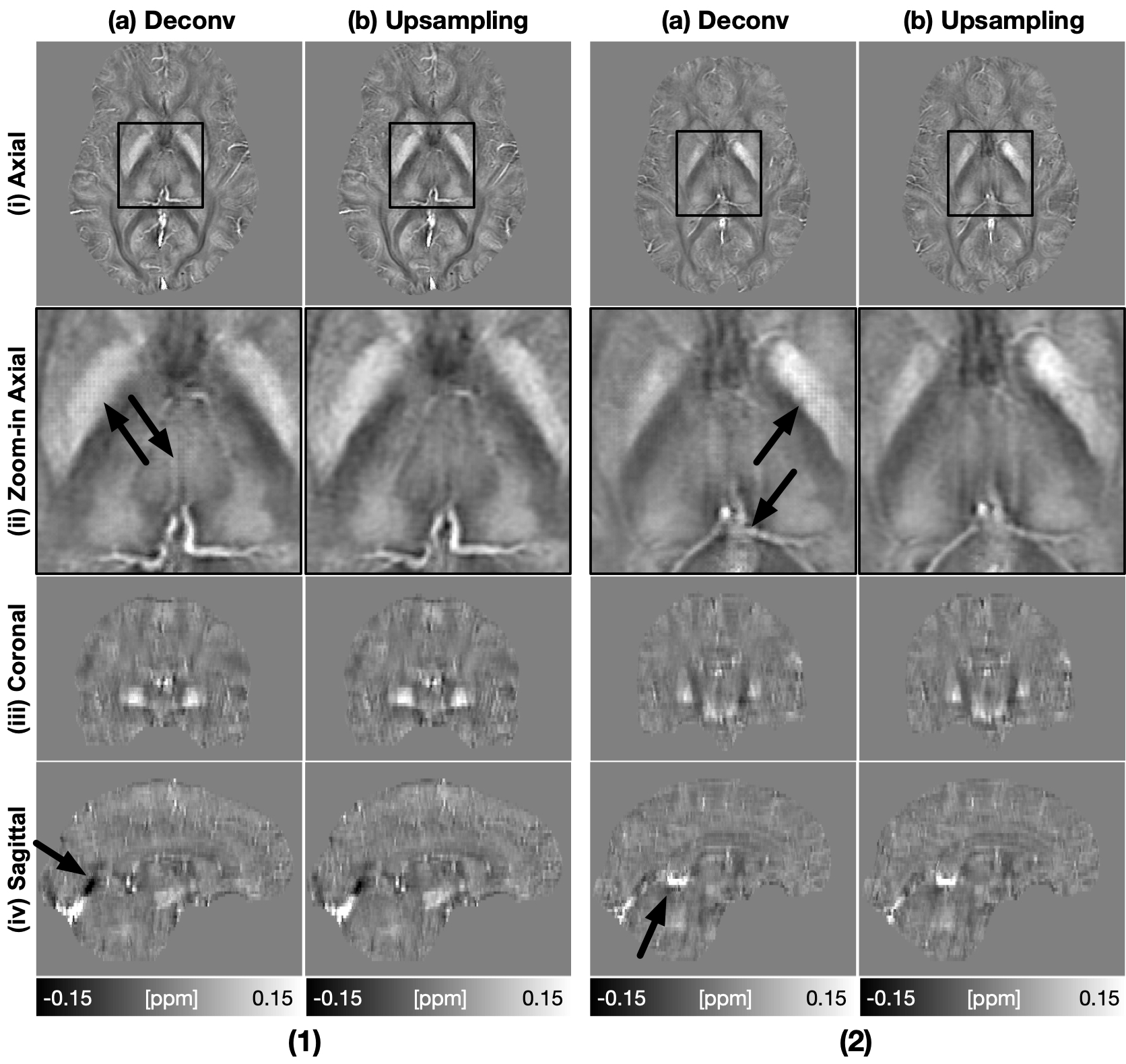}
\vspace{-10pt}
\caption{Comparison of QSM results of 2 single-orientation datasets. uQSM using deconvolution (a) showed checkboard artifacts in zoom-in axial plane (a, ii, black arrows).}
\label{fig:uQSM_upSamplingH2H}
\vspace{-30pt}
\end{center}
\end{figure}

\textbf{Deconvolution and Checkboard Artifacts}
In uQSM, we used upsampling layers rather than deconvolutional layers to reduce the checkerboard artifacts \cite{odena2016deconvolution} in QSM results. Here we trained two networks on the single-orientation datasets, one with upsampling and the other with deconvolution in the decoding path. Fig.\ref{fig:uQSM_upSamplingH2H} compared QSM images of 2 single-orientation datasets reconstructed using deconvolution-based network and upsampling-based network. Deconvolution-based network produced QSM with checkboard artifacts in the zoom-in axial plane (a, ii, black arrows).

\textbf{Effects of $L_{TV}$ and Data Consistency Losses} We used the multi-orientation datasets to investigate the effects of $L_{TV}$ and three data consistency losses - (1) linear dipole inversion (LDI), $L_{LDI} = \left \|m(d\ast \chi  -  y) \right \|_{2}$, (2) weighted linear dipole inversion (WLDI), $L_{WLDI} = \left \|W m (e^{i d\ast \chi}  - e^{i y}) \right \|_{2}$, (3) weighted nonlinear dipole inversion (NDI), $L_{NDI} = \left \| W m (d\ast \chi  -  y) \right \|_{2}$.

\begin{table}[H]
\centering
\vspace{-15pt}
\caption{\label{tab:uQSM_Loss} Means and standard deviations of quantitative performance metrics of uQSM using different loss function on 9 multi-orientation datasets.}
\vspace{-5pt}
\begin{tabular}{cccccc}
\hline
\multicolumn{1}{|c}{} & \multicolumn{1}{|c}{pSNR (dB)} & \multicolumn{1}{|c}{NRMSE ($\%$)} &\multicolumn{1}{|c}{HFEN ($\%$)} &\multicolumn{1}{|c|}{SSIM (0-1)}\\
\hline 
\multicolumn{1}{|c}{$L_{NDI}$} & \multicolumn{1}{|c}{$43.8\pm0.5$} & \multicolumn{1}{|c}{$87.4\pm6.8$} & \multicolumn{1}{|c}{$70.5\pm5.7$} & \multicolumn{1}{|c|}{{$0.848\pm0.022$}}\\
\hline
\multicolumn{1}{|c}{$L_{LDI} + \lambda L_{TV}$} & \multicolumn{1}{|c}{$44.1\pm0.5$} & \multicolumn{1}{|c}{$84.9\pm5.9$} & \multicolumn{1}{|c}{$73.4\pm5.7$} & \multicolumn{1}{|c|}{{$0.879\pm0.013$}}\\
\hline
\multicolumn{1}{|c}{$L_{LWDI} + \lambda L_{TV}$} & \multicolumn{1}{|c}{$45.0\pm0.5$} & \multicolumn{1}{|c}{$75.9\pm5.4$} & \multicolumn{1}{|c}{$67.1\pm5.5$} & \multicolumn{1}{|c|}{{$0.888\pm0.015$}}\\
\hline
\multicolumn{1}{|c}{$L_{NDI} + \lambda L_{TV}$} & \multicolumn{1}{|c}{\textbf{45.6$\pm$0.4}} & \multicolumn{1}{|c}{\textbf{71.4$\pm$5.0}} & \multicolumn{1}{|c}{\textbf{62.8$\pm$5.0}} & \multicolumn{1}{|c|}{\textbf{0.890$\pm$0.015}}\\
\hline 
\vspace{-35pt}
\end{tabular}
\begin{flushleft}
\end{flushleft}
\end{table}

\begin{figure}[H]
\begin{center}
\vspace{-10pt}
\includegraphics[width=0.8\textwidth]{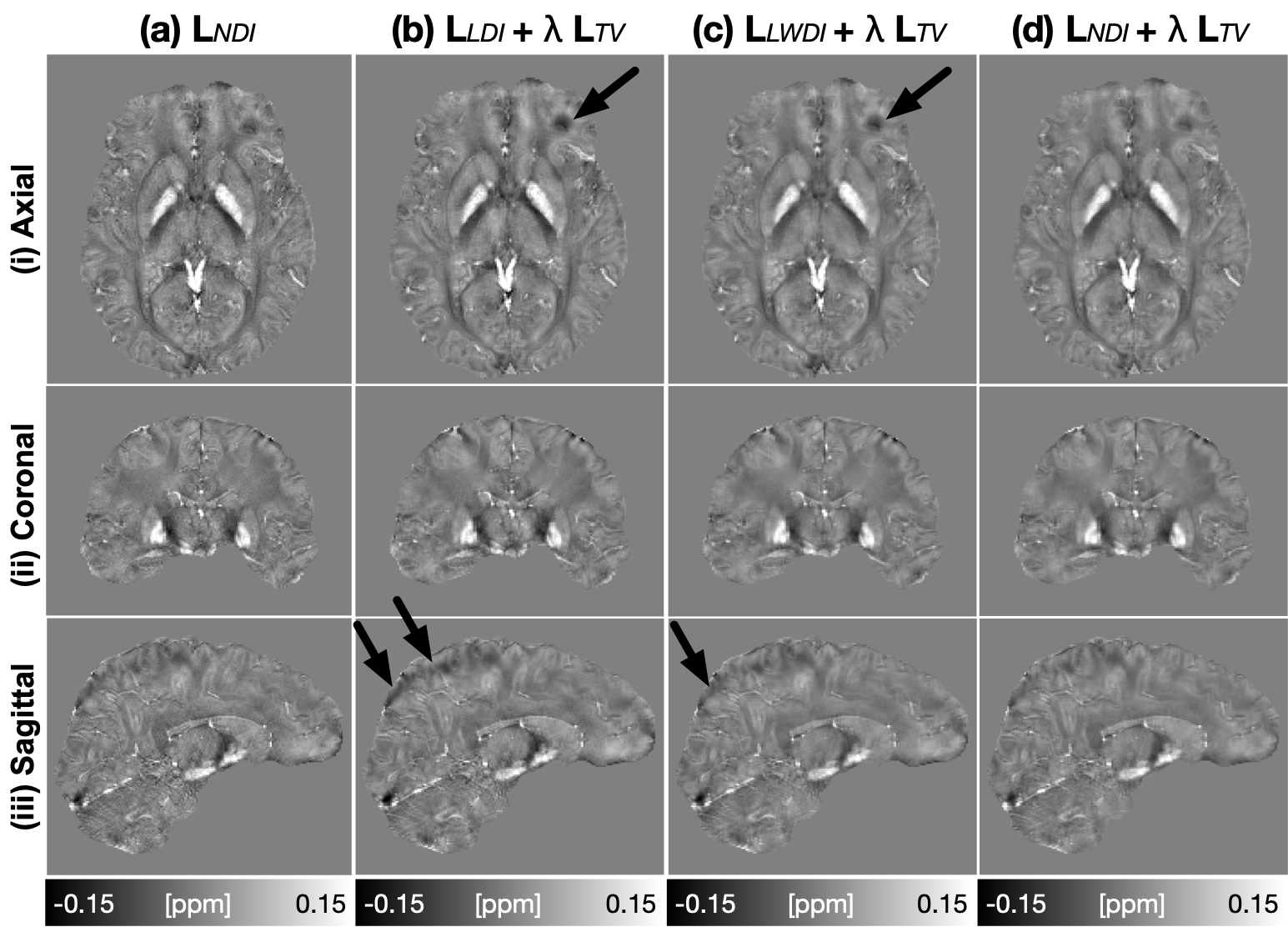}
\vspace{-5pt}
\caption{Comparison of uQSM results using different loss functions of one multi-orientation datasets.}
\label{fig:uQSM_loss}
\vspace{-20pt}
\end{center}
\end{figure}

Table.\ref{tab:uQSM_Loss} summarized quantitative metrics on 9 multi-orientation datasets with COSMOS map as a reference. uQSM using $L_{NDI} + \lambda L_{TV}$ achieved the best metric scores. Fig.\ref{fig:uQSM_loss} displayed QSM images. Without $L_{TV}$, the QSM showed high level of noise (a). Using $L_{LDI}$ and $L_{WLDI}$ as data consistency loss, the QSM estimates displayed black shading artifacts (b-c, i-iii, black arrows), while $L_{NDI}$ was capable of suppressing these artifacts.

\section{Discussion and Conclusion}
In this work, a model-based DL method for QSM dipole deconvolution was proposed. Without accessing to QSM labels during training, uQSM learned to perform dipole inversion through the physical model. 

From quantitative evaluation on multi-orientation QSM datasets, uQSM outperformed TKD, TV-FANSI, MEDI, and DIP in pSNR, RMSE, and HFEN, with COSMOS map as a reference. The visual assessment demonstrated that uQSM preserved image details well and showed invisible image artifacts. When using single-orientation datasets for qualitative assessment, uQSM results showed better image quality than conventional non-DL methods and DIP. Though DIP is unsupervised and does not require training data for prior training, it needs long iteration times for each dataset and early-stopping to avoid overfitting. In addition, the upsampling used in uQSM network can avoid the checkerboard artifacts in the QSM estimates. $L_{TV}$ enable to denosing the QSM outputs and preserve the edge information. $L_{NDI}$ as data consistency loss can improve the image quality of uQSM than $L_{LDI}$ and $L_{LWDI}$.

Future work can apply more sophisticated models \cite{schweser2018quantitative} in uQSM. In addition, uQSM is still affected by the performance of background field removal methods. It is necessary to investigate the effects of background field removal on susceptibility quantification or perform DL-based single-step QSM reconstruction. 

\section*{Acknowledgement}

We thank Professor Jongho Lee for sharing the multi-orientation QSM datasets.

\bibliographystyle{splncs04}
\bibliography{references}

\end{document}


\title{--------- Supplemental Document ---------  \protect\\                
       Model-based Learning for Quantitative Susceptibility Mapping}
\titlerunning{Model-based Learning for QSM Reconstruction}
%

\author{{Juan Liu\inst{1,2}} \and
Kevin M. Koch\inst{1,2,3}}
\authorrunning{J. Liu, K.M. Koch}

\institute{Center for Imaging Research, Medical College of Wisconsin, Milwaukee, WI, USA \and
Department of Biomedical Engineering, Marquette University and Medical College of Wisconsin, Milwaukee, WI, USA \and
Department of Radiology, Medical College of Wisconsin, Milwaukee, WI, USA}

\maketitle  

\section{Patch-based Training Scheme}
In view of the limited available QSM data and the hardware constraints, uQSM was trained using patch-based neural networks. Patch-based neural networks with small patch size might lack enough contextual information required for non-local susceptibility estimation, which will cause larger susceptibility quantification errors. Though larger patches even full image network training may increase the context and produce more accurate susceptibility estimates, this will significantly increase the amount of memory and computation requirements. Especially, we used 3D convolutional neural networks for QSM reconstruction, large patch-based neural networks or full image neural networks consume prohibitively large amounts of memory, causing computational burden. In addition, patch-based neural networks can be trained with more training patches, which alleviates the limited available training data problem. In uQSM, the model-based data consistency loss was used, which could introduce errors in susceptibility estimates when using patch-based neural networks. Therefore, it is important to know how patch-based neural networks performs in model-based QSM learning and how to choose the appropriate patch size. 

In order to investigate the effects of patch size and supervision in QSM, we used synthetic data to train different patch-based neural networks in supervised and unsupervised way. The synthetic training data with matrix size (160, 160, 160) and pixel resolution 1.0x1.0x1.0 mm$^3$ were generated using a single COSMOS dataset and data augmentation techniques in this experiment. The single COSMOS dataset of 2016 QSM challenge was used to generate the training data. In addition, we applied elastic transform, image contrast change, and superimposing high susceptibility sources inside the brain to augment the COSMOS map. The local fields were then calculated using the dipole convolution without adding image noise. Since it is difficult to get large amounts of datasets in medical imaging, only 1000 synthetic datasets were generated for training. 

5 neural networks were trained with patch size (48,48,48), (64, 64, 64), (96, 96, 96), (128, 128, 128) and (160, 160, 160) in a supervised way with L2 loss, with the same neural network architecture aforementioned. In addition, 5 neural networks were trained with patch size (48,48,48), (64, 64, 64), (96, 96, 96), (128, 128, 128) and (160, 160, 160) in an unsupervised way with the loss $Loss = \left \| m (d\ast \chi  - f) \right \|_{2}$. During inference, the trained models took the full local fields and brain masks to get QSM images, since the convolutional layer conducted their operation sequentially regardless of the input matrix size.. 

For quantitative evaluation, 100 testing data were generated as the same as training data. Quantitative metrics, peak signal-to-noise ratio (pSNR), normalized root mean squared error (NRMSE), high-frequency error norm (HFEN), and structure similarity index (SSIM), were used for evaluation in comparison with the ground truth.

\begin{table}[H]
\centering
\caption{\label{tab:uQSM_Patch_Supervised} Means and standard deviations of the quantitative performance metrics, pSNR, NRMSE, HFEN, and SSIM of 5 DL models in a supervised way on 100 testing data.}
\vspace{0in}
\begin{tabular}{cccccc}
\hline
\multicolumn{1}{|c}{} & \multicolumn{1}{|c}{pSNR(dB)} & \multicolumn{1}{|c}{NRMSE($\%$)} &\multicolumn{1}{|c}{HFEN($\%$)} &\multicolumn{1}{|c|}{SSIM (0-1)}\\
\hline
\multicolumn{1}{|c}{QSM (48)} & \multicolumn{1}{|c}{$49.0\pm1.3$} & \multicolumn{1}{|c}{$28.1\pm2.5$} & \multicolumn{1}{|c}{$30.8\pm2.3$} & \multicolumn{1}{|c|}{$0.964\pm0.021$}\\
\hline
\multicolumn{1}{|c}{QSM (64)} & \multicolumn{1}{|c}{$50.0\pm1.3$} & \multicolumn{1}{|c}{$25.4\pm4.2$} & \multicolumn{1}{|c}{$27.8\pm2.1$} & \multicolumn{1}{|c|}{$0.968\pm0.019$}\\
\hline
\multicolumn{1}{|c}{QSM (96)} & \multicolumn{1}{|c}{\bf{51.1$\pm$1.4}} & \multicolumn{1}{|c}{\bf{22.7$\pm$7.6}} & \multicolumn{1}{|c}{\bf{23.5$\pm$1.8}} & \multicolumn{1}{|c|}{$0.975\pm0.015$}\\
\hline
\multicolumn{1}{|c}{QSM (128)} & \multicolumn{1}{|c}{$51.1\pm1.5$} & \multicolumn{1}{|c}{$23.2\pm10.9$} & \multicolumn{1}{|c}{$23.6\pm2.9$} & \multicolumn{1}{|c|}{\bf{0.976$\pm$0.014}}\\
\hline
\multicolumn{1}{|c}{QSM (160)} & \multicolumn{1}{|c}{$50.4\pm1.4$} & \multicolumn{1}{|c}{$25.3\pm13.6$} & \multicolumn{1}{|c}{$26.0\pm1.9$} & \multicolumn{1}{|c|}{$0.970\pm0.018$}\\
\hline
\end{tabular}
\begin{flushleft}
\end{flushleft}
\end{table}

Table.\ref{tab:uQSM_Patch_Supervised} summarized the quantitative evaluation on 100 testing data from 5 supervised neural networks. Networks trained with larger patch size (96, 96, 96) and (128, 128, 128) achieved better quantitative scores than smaller patch size (48, 48, 48) and (64, 64, 64). The network trained with full image size (160, 160, 160) achieved subtle poorer metric scores than patch-based neural network (96, 96, 96) and (128, 28, 128), which may be due to the limited training data for full image training. 

\begin{figure}[H]
\begin{center}
\vspace{-15pt}
\includegraphics[width=\textwidth]{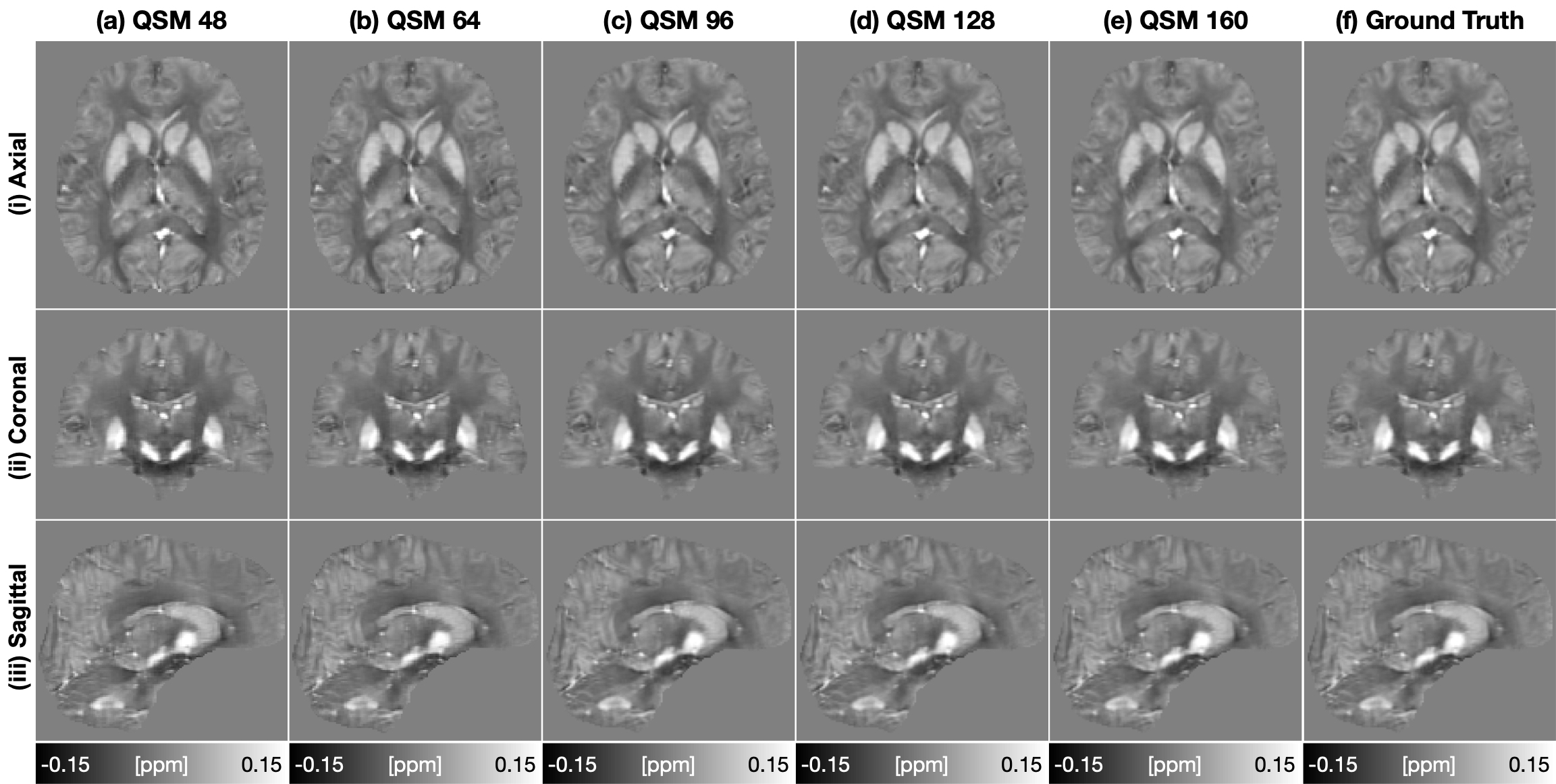}
\caption{Comparison of QSM results (a-e) from 5 supervised networks on a testing data.}
\label{fig:uQSM_Patch_Supervised}
\vspace{-15pt}
\end{center}
\end{figure}

\begin{figure}[H]
\begin{center}
\vspace{-15pt}
\includegraphics[width=\textwidth]{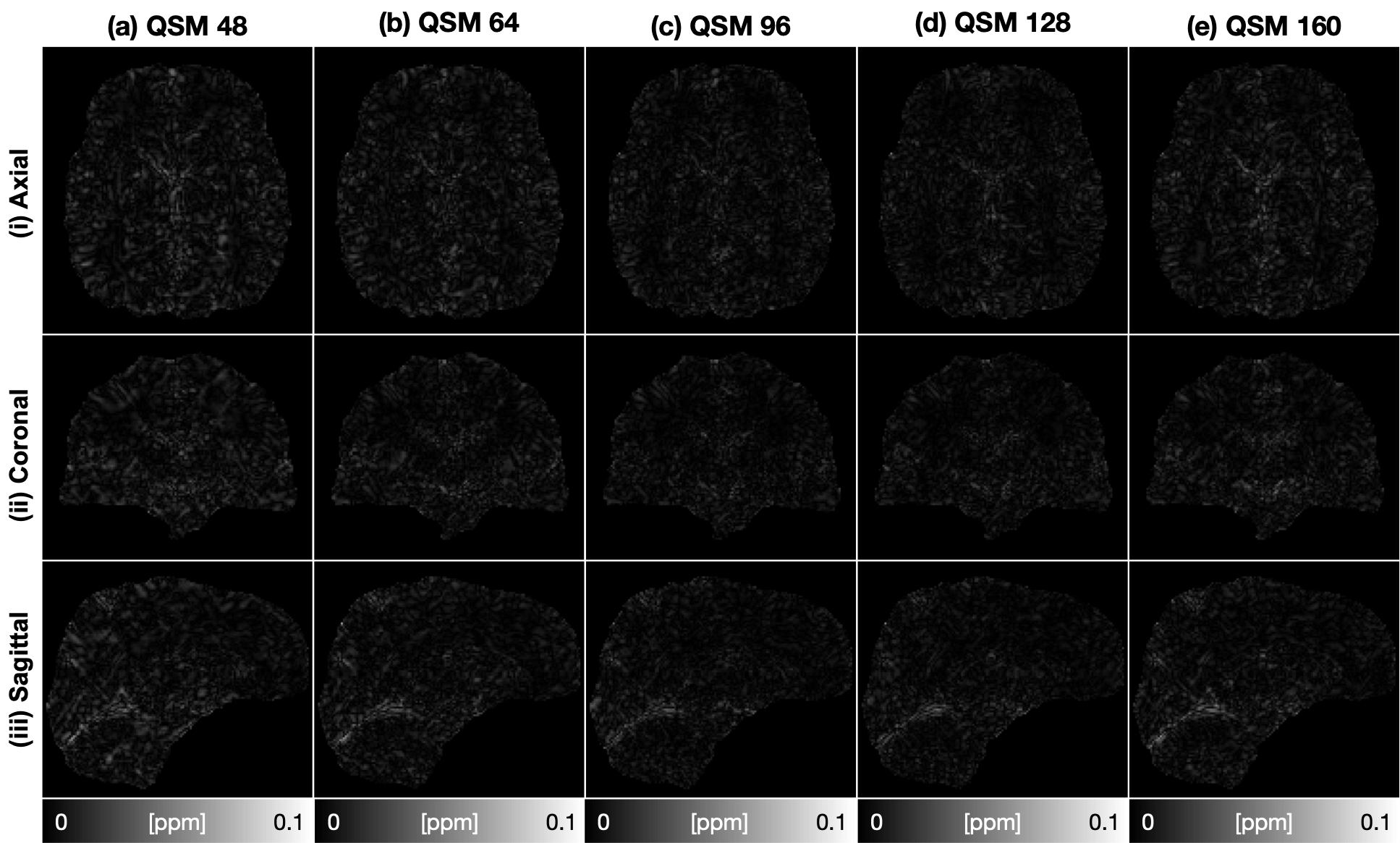}
\caption{Residual error maps of Fig.\ref{fig:uQSM_Patch_Supervised} on a testing data from 5 supervised networks. All maps showed small errors.}
\label{fig:uQSM_Patch_Supervised_ErrMap}
\vspace{-15pt}
\end{center}
\end{figure}

Fig.\ref{fig:uQSM_Patch_Supervised} displayed QSM results from a testing data from 5 supervised networks. Fig.\ref{fig:uQSM_Patch_Supervised_ErrMap} showed the corresponding residual error maps. All supervised neural networks produced QSM images with comparable image quality and small residual errors.

Table.\ref{tab:uQSM_Patch_Unsupervised} summarized the quantitative evaluation on 100 testing data from 5 model-based neural networks. Networks trained with larger patch size achieved better quantitative scores than smaller patch size. Patch-based neural networks with patch size (48, 48, 48) and (64, 64, 64) have poor quantitative scores in pSNR, RMSE, HFEN, and SSIM. Patch-based neural networks with patch size (96, 96, 96) achieved comparable metric scores with (128, 128, 128) and (160, 160, 160). Patch-based neural network (128, 128, 128) achieved the best in all quantitative scores than other networks. Compared with Table.\ref{tab:uQSM_Patch_Supervised}, supervised models achieved better quantitative scores than unsupervised models.

\begin{table}[H]
\centering
\caption{\label{tab:uQSM_Patch_Unsupervised} Means and standard deviations of the quantitative performance metrics, pSNR, NRMSE, HFEN, and SSIM of 5 unsupervised models on 100 testing data.}
\vspace{0in}
\begin{tabular}{cccccc}
\hline
\multicolumn{1}{|c}{} & \multicolumn{1}{|c}{pSNR(dB)} & \multicolumn{1}{|c}{NRMSE($\%$)} &\multicolumn{1}{|c}{HFEN($\%$)} &\multicolumn{1}{|c|}{SSIM (0-1)}\\
\hline 
\multicolumn{1}{|c}{uQSM (48)} & \multicolumn{1}{|c}{$42.2\pm0.8$} & \multicolumn{1}{|c}{$61.1\pm3.8$} & \multicolumn{1}{|c}{$50.1\pm3.3$} & \multicolumn{1}{|c|}{$0.935\pm0.037$}\\
\hline
\multicolumn{1}{|c}{uQSM (64)} & \multicolumn{1}{|c}{$43.7\pm0.8$} & \multicolumn{1}{|c}{$51.7\pm2.9$} & \multicolumn{1}{|c}{$44.0\pm2.8$} & \multicolumn{1}{|c|}{$0.946\pm0.031$}\\
\hline
\multicolumn{1}{|c}{uQSM (96)} & \multicolumn{1}{|c}{$47.5\pm1.2$} & \multicolumn{1}{|c}{$33.2\pm1.3$} & \multicolumn{1}{|c}{$35.7\pm2.4$} & \multicolumn{1}{|c|}{$0.960\pm0.023$}\\
\hline
\multicolumn{1}{|c}{uQSM (128)} & \multicolumn{1}{|c}{\bf{48.6$\pm$1.3}} & \multicolumn{1}{|c}{\bf{29.5$\pm$2.1}} & \multicolumn{1}{|c}{\bf{35.2$\pm$2.4}} & \multicolumn{1}{|c|}{\bf{0.961$\pm$0.022}}\\
\hline
\multicolumn{1}{|c}{uQSM (160)} & \multicolumn{1}{|c}{$48.5\pm1.3$} & \multicolumn{1}{|c}{$29.7\pm1.7$} & \multicolumn{1}{|c}{$36.4\pm2.3$} & \multicolumn{1}{|c|}{$0.959\pm0.023$}\\
\hline
\end{tabular}
\begin{flushleft}
\end{flushleft}
\end{table}

\begin{figure}[H]
\begin{center}
\vspace{-15pt}
\includegraphics[width=\textwidth]{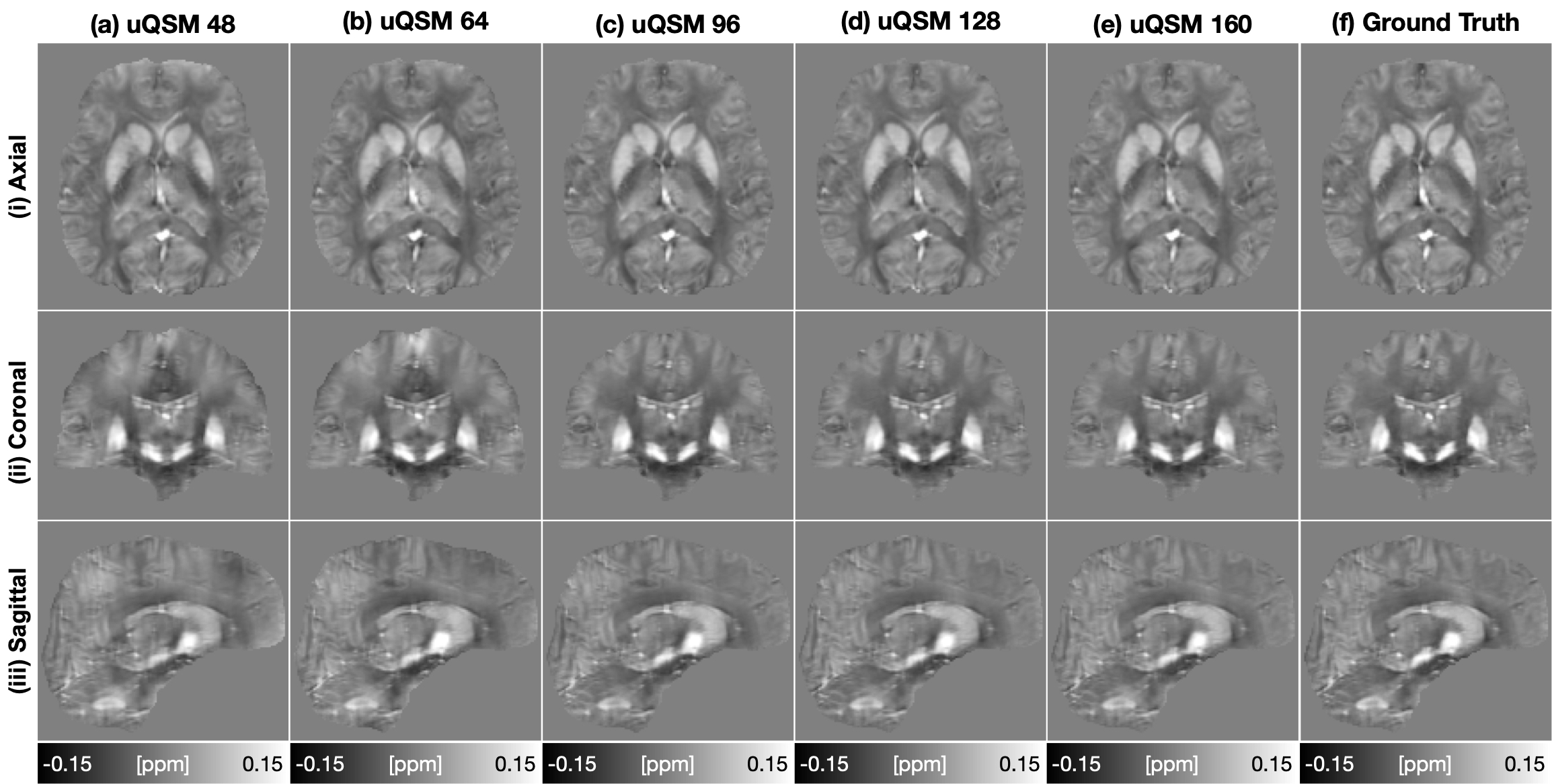}
\caption{Comparison of QSM results (a-e) from 5 unsupervised networks on a testing data.}
\label{fig:uQSM_Patch_Unsupervised}
\vspace{-15pt}
\end{center}
\end{figure}

\begin{figure}[H]
\begin{center}
\vspace{-15pt}
\includegraphics[width=\textwidth]{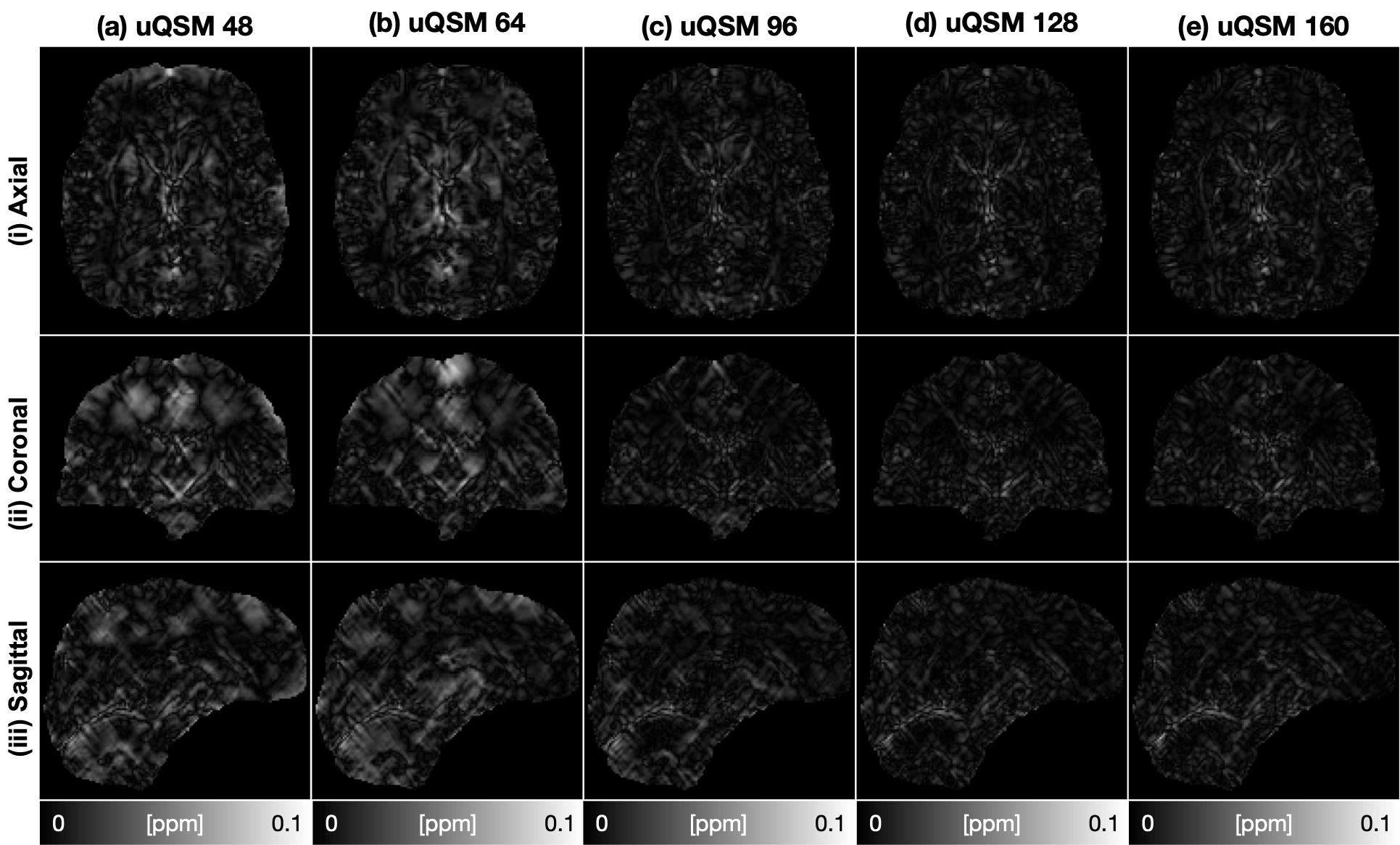}
\caption{Residual error maps of Fig.\ref{fig:uQSM_Patch_Unsupervised} on a testing data from 5 unsupervised networks. Networks using patch size (48, 48, 48) and (64, 64, 64) showed large residual errors than networks using (96, 96, 96), (128, 128, 128) and (160, 160, 160).}
\label{fig:uQSM_Patch_Unsupervised_ErrMap}
\vspace{-15pt}
\end{center}
\end{figure}

Fig.\ref{fig:uQSM_Patch_Unsupervised} displayed QSM results from a testing data from 5 unsupervised networks and Fig.\ref{fig:uQSM_Patch_Unsupervised_ErrMap} showed the corresponding residual error maps. Compared with supervised methods, unsupervised networks perform worse. Unsupervised networks with patch size (48, 48, 48) and (64, 64, 64) got poor quantitative metrics scores and large residual errors than models with patch size (96, 96, 96), (128, 128, 128) and (160, 160, 160). Network with patch size (96, 96, 96) achieved comparable quantitative metrics scores with (128, 128, 128) and (160, 160, 160). Therefore, it is preferred to apply patch-based neural networks with large patch size for unsupervised QSM learning.

\subsection{Effects of Patch Size on Multi-orientation Datasets}
We used 5 patch-based neural networks with patch size (48, 48, 48), (64, 64, 64), (96, 96, 96), (128, 128, 128) and (160, 160, 160) on multi-orientation datasets. Leave-one-out cross validation was used. The 5 nerual networks were trained with the same network architecture described in the paper and hyperparameters. During inference, the trained models took the full local fields and brain masks to get the QSM images.

Table.\ref{tab:uQSM_patch_QSMdata} summarized quantitative metrics of 5 patch-based neural networks on multi-orientation datasets using the COSMOS map as a reference. The results showed that the neural networks trained with different patch sizes achieved similar metric scores. uQSM using patch size (64, 64, 64) and (96, 96, 96) achieved slightly better in pSNR, NRMSE, HFEN, and SSIM than uQSM using patch size (48, 48, 48), (128, 128, 128), and (160, 160, 160).

\begin{table}[H]
\centering
\caption{\label{tab:uQSM_patch_QSMdata} Means and standard deviations of the quantitative performance metric, pSNR, NRMSE, HFEN, and SSIM of uQSM with different patch size on 9 multi-orientation datasets.}
\vspace{0in}
\begin{tabular}{cccccc}
\hline
\multicolumn{1}{|c}{} & \multicolumn{1}{|c}{pSNR (dB)} & \multicolumn{1}{|c}{NRMSE ($\%$)} &\multicolumn{1}{|c}{HFEN ($\%$)} &\multicolumn{1}{|c|}{SSIM (0-1)}\\
\hline 
\multicolumn{1}{|c}{uQSM 48} & \multicolumn{1}{|c}{$45.3\pm0.5$} & \multicolumn{1}{|c}{$73.8\pm5.0$} & \multicolumn{1}{|c}{$61.7\pm4.7$} & \multicolumn{1}{|c|}{$0.887\pm0.015$}\\
\hline
\multicolumn{1}{|c}{uQSM 64} & \multicolumn{1}{|c}{$45.7\pm0.5$} & \multicolumn{1}{|c}{\bf{70.0$\pm$5.3}} & \multicolumn{1}{|c}{\bf{61.1$\pm$5.5}} & \multicolumn{1}{|c|}{\bf{0.890$\pm$0.015}}\\
\hline
\multicolumn{1}{|c}{uQSM 96} & \multicolumn{1}{|c}{\bf{45.7$\pm$0.4}} & \multicolumn{1}{|c}{$71.4\pm5.0$} & \multicolumn{1}{|c}{$62.8\pm5.0$} & \multicolumn{1}{|c|}{\bf{0.890$\pm$0.015}}\\
\hline
\multicolumn{1}{|c}{uQSM 128} & \multicolumn{1}{|c}{$45.3\pm0.4$} & \multicolumn{1}{|c}{$73.8\pm5.9$} & \multicolumn{1}{|c}{$63.8\pm5.8$} & \multicolumn{1}{|c|}{$0.888\pm0.015$}\\
\hline
\multicolumn{1}{|c}{uQSM 160} & \multicolumn{1}{|c}{$45.4\pm0.5$} & \multicolumn{1}{|c}{$73.1\pm5.3$} & \multicolumn{1}{|c}{$62.8\pm5.1$} & \multicolumn{1}{|c|}{$0.886\pm0.014$}\\
\hline
\vspace{-15pt}
\end{tabular}
\begin{flushleft}
\end{flushleft}
\end{table}

\begin{figure}[h]
\begin{center}
\vspace{-15pt}
\includegraphics[width=\textwidth]{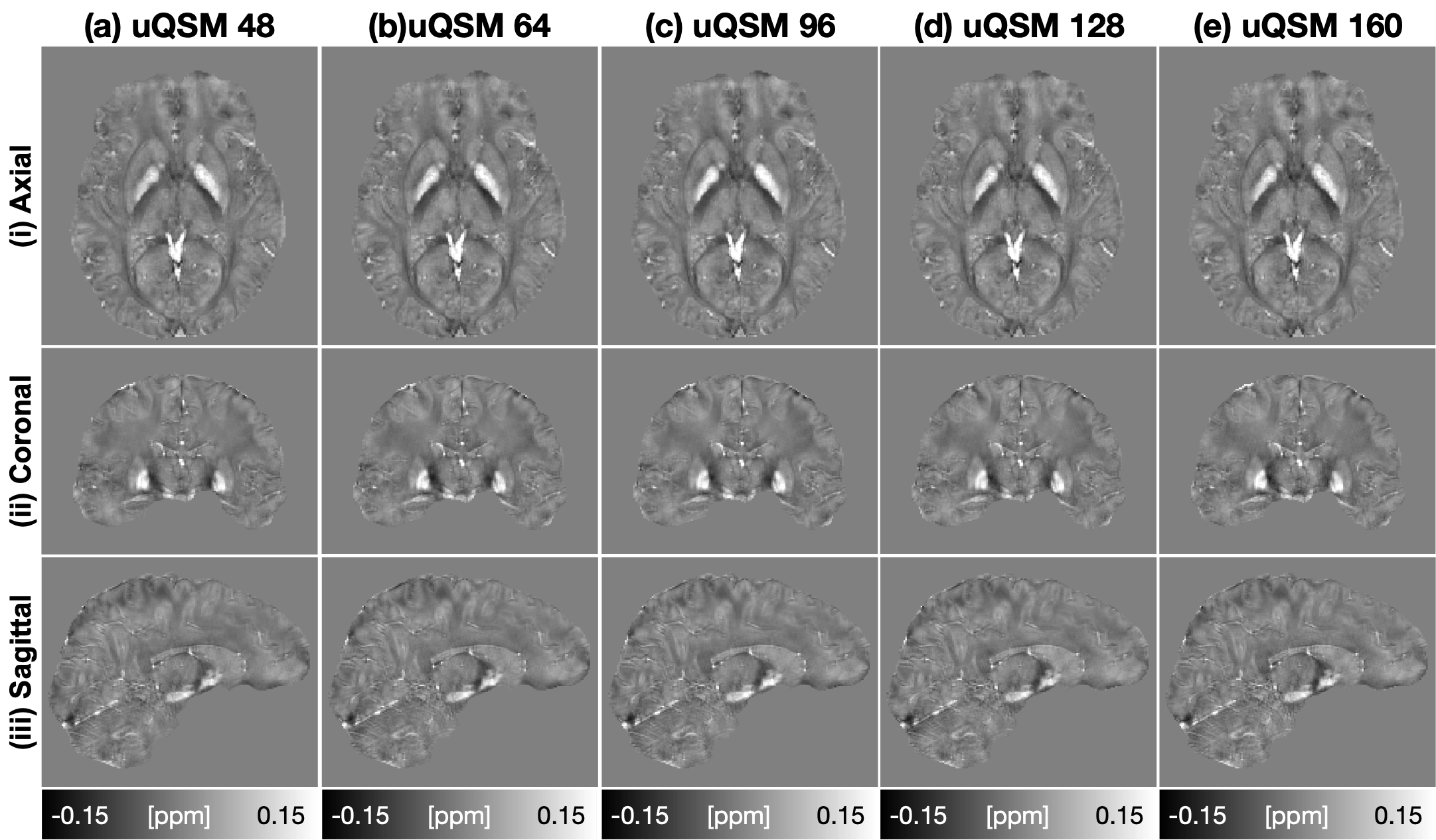}
\caption{Comparison of QSM results on a multi-orientation datasets from 5 patch-based neural networks. All show similar image quality.}
\label{fig:uQSM_Patch_QSMdata}
\vspace{-15pt}
\end{center}
\end{figure}

Fig.\ref{fig:uQSM_Patch_QSMdata} showed the QSM images of a multi-orientation datasets from 5 patch-based neural networks. Visually comparison showed that they have very similar images. Quantitative and qualitative comparison demonstrated that patch size may have small effects on the QSM estimates. 

Though the multi-orientation datasets showed patch size does not have high effects on the QSM estimates, it is more proffered to train uQSM with large patch size for non-local susceptibility estimation. In the multi-orientation datasets, the COSMOS map as a reference contains errors from background field removal and image registration procedures, which cannot guarantee a `ground-truth'. In addition, the quantitative metrics pSNR, NRMSE, HFEN, and SSIM are global error metrics aiming to summarize the mismatch against a reference image in a single number. As it pointed out by 2016 QSM reconstruction challenge report, QSM results with higher quantitative metrics cannot truly reflect the QSM image quality. For example, algorithms yielded highly over‐regularized smooth QSM images which achieved better RMSE scores. Future work is necessary to investigate how to choose the proper patch size for QSM datasets with different image resolution.


\section{Comparison with DIP}
DIP used the same neural network architecture as uQSM. DIP was performed for each individual dataset without prior network training. By taking a full local field map and a brain mask as the inputs, DIP used the same loss function as uQSM to update neural network parameters and the QSM output. To avoid overfitting which can result in artifacts in the reconstructed QSM images, DIP was early stopping after 200 iterations to get QSM estimates. 

\begin{figure}[H]
\begin{center}
\includegraphics[width=\textwidth]{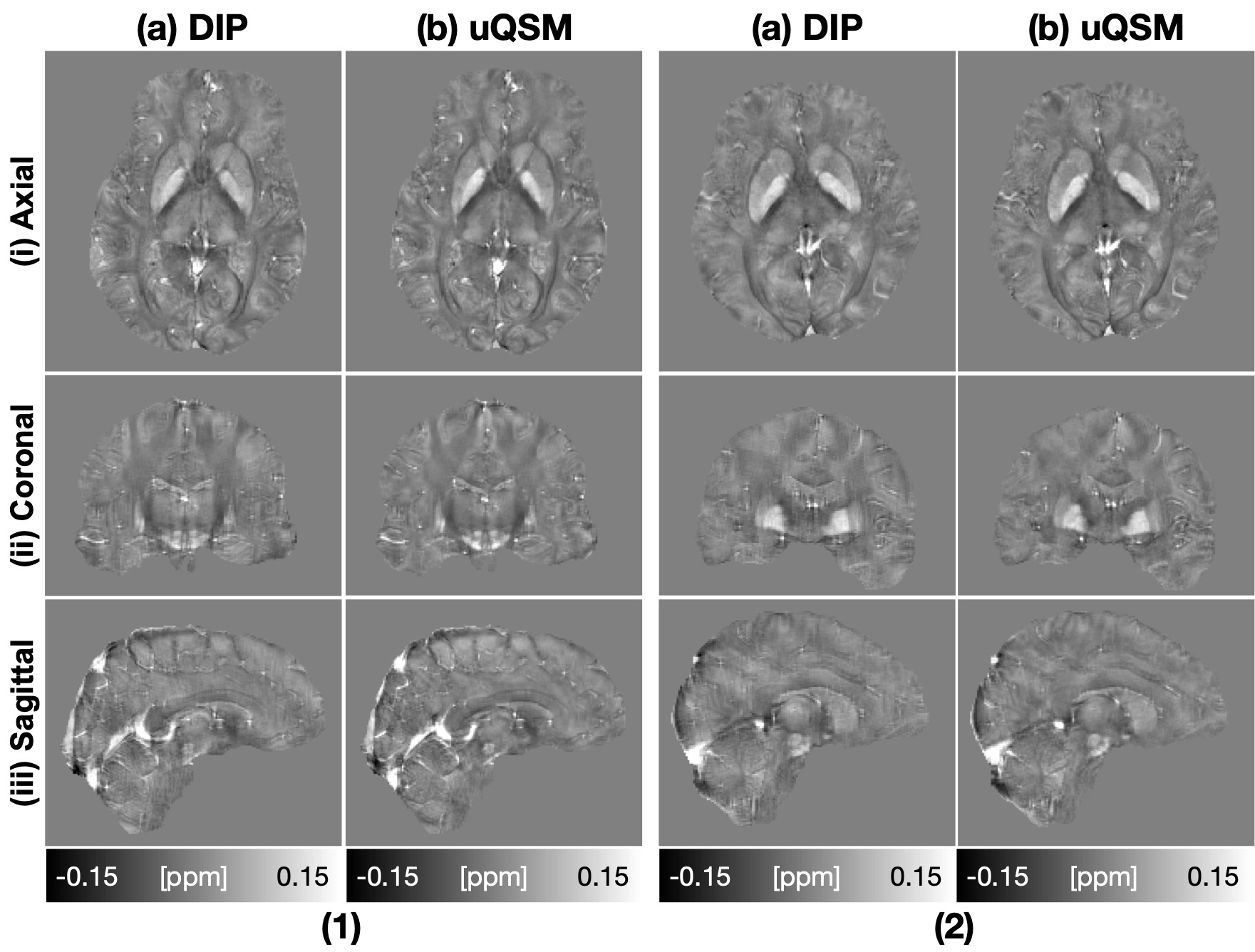}
\caption{Comparison of QSM results from two multi-orientation datasets reconstructed by DIP and uQSM. DIP and uQSM images showed visually similar.}
\label{fig:uQSM_DIP_QSMdata}
\end{center}
\end{figure}

Fig.\ref{fig:uQSM_DIP_QSMdata} displayed QSM images from two multi-orientation datasets reconstructed by DIP and uQSM. DIP and uQSM results showed comparable image quality.

\begin{figure}[h]
\begin{center}
\vspace{-10pt}
\includegraphics[width=.9\textwidth]{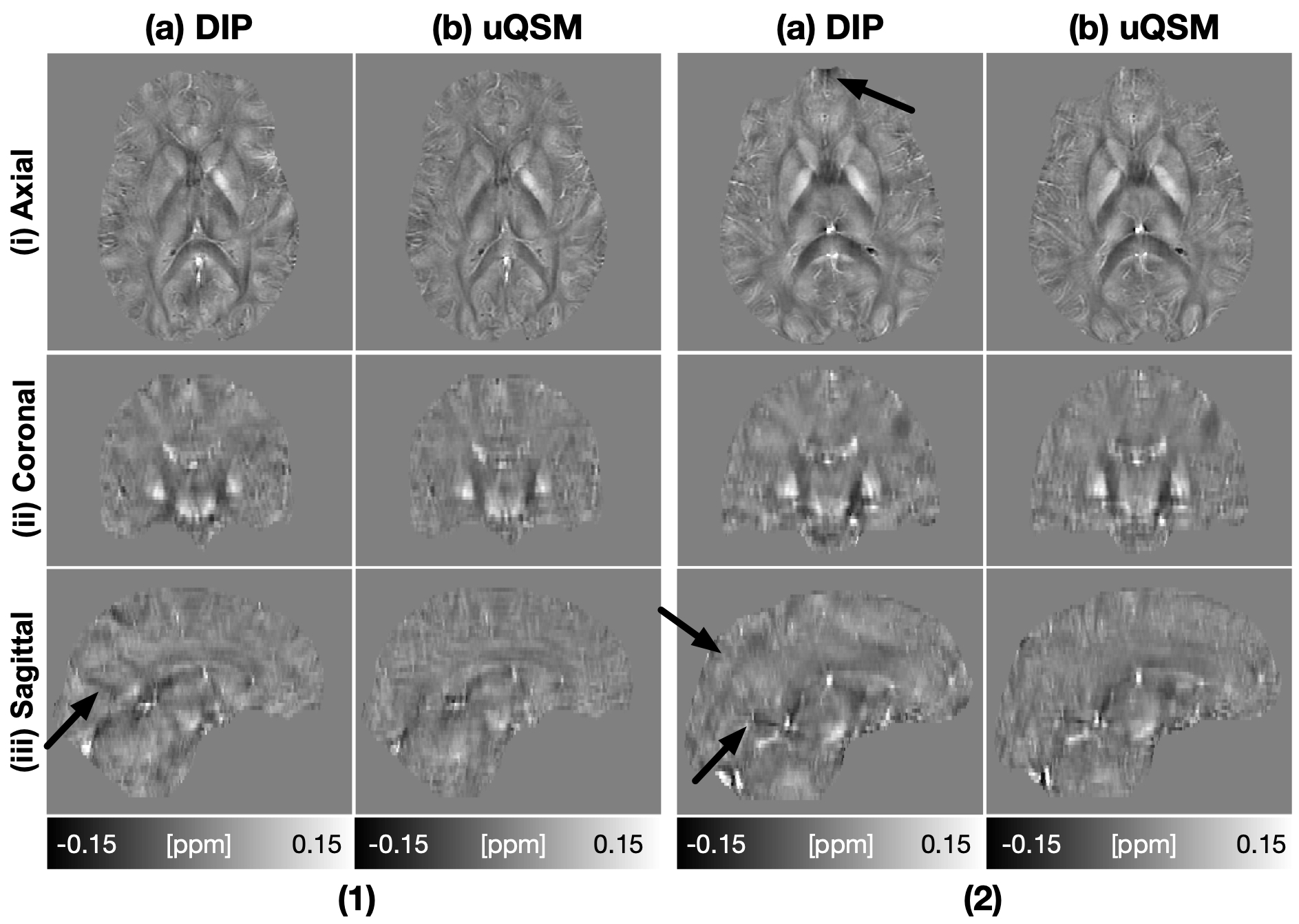}
\caption{Comparison of QSM results from two single-orientation datasets reconstructed by DIP and uQSM. DIP results had visible streaking and shading artifacts (a, black arrows), while uQSM images displayed invisible image artifacts.}
\label{fig:uQSM_DIP_H2H}
\vspace{-20pt}
\end{center}
\end{figure}

\begin{figure}[H]
\vspace{-10pt}
\begin{center}
\includegraphics[width=.9\textwidth]{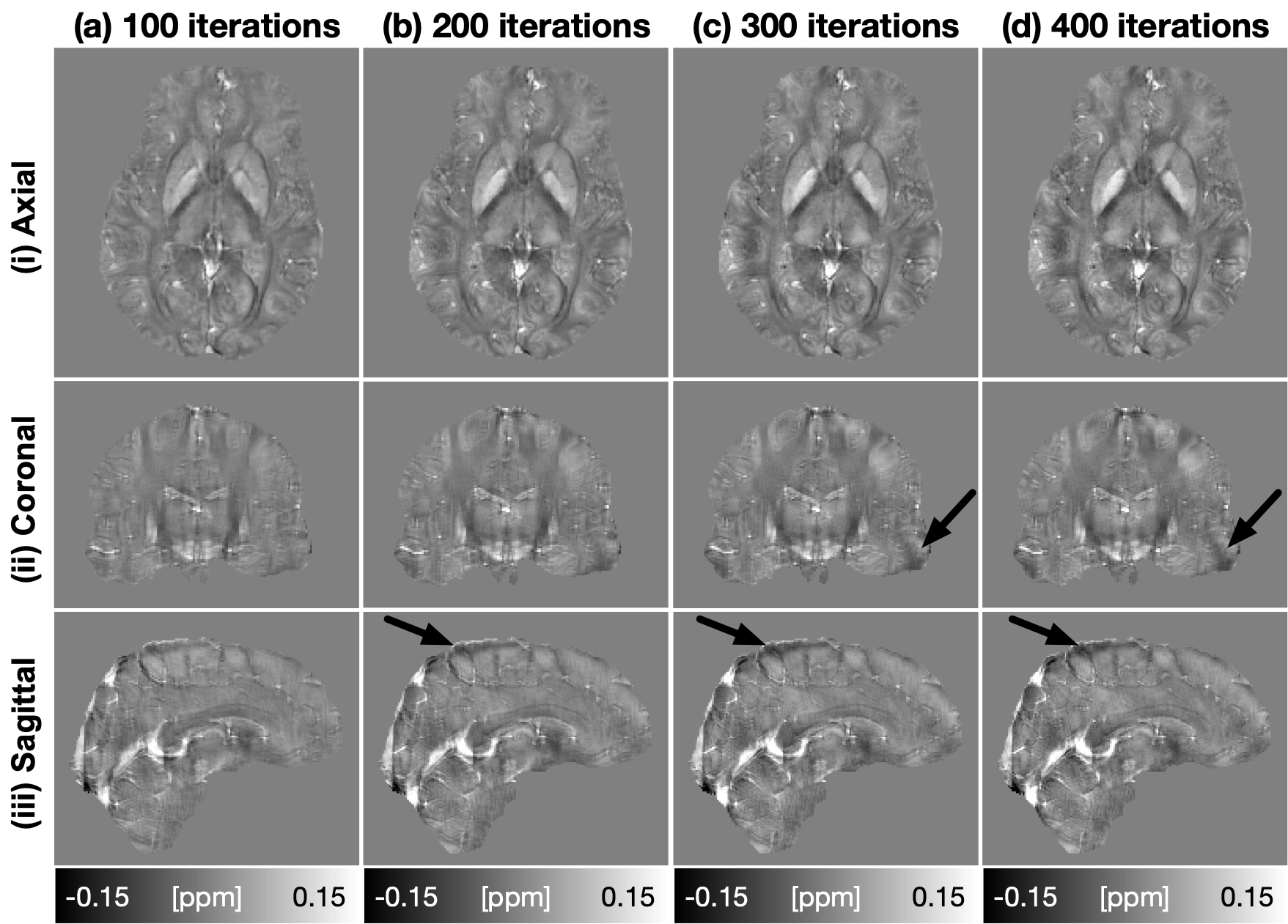}
\caption{Comparison of DIP results of a multi-orientation dataset. DIP results of 100 iterations (a) showed inferior image contrast yet invisible image artifacts. Streaking and shading artifacts became more apparent in DIP results of 200, 300, and 400 iterations (b-d, black arrows).}
\label{fig:uQSM_DIP_QSMdata_Steps_1}
\end{center}
\vspace{-20pt}
\end{figure}

\begin{figure}[H]
\vspace{-10pt}
\begin{center}
\includegraphics[width=\textwidth]{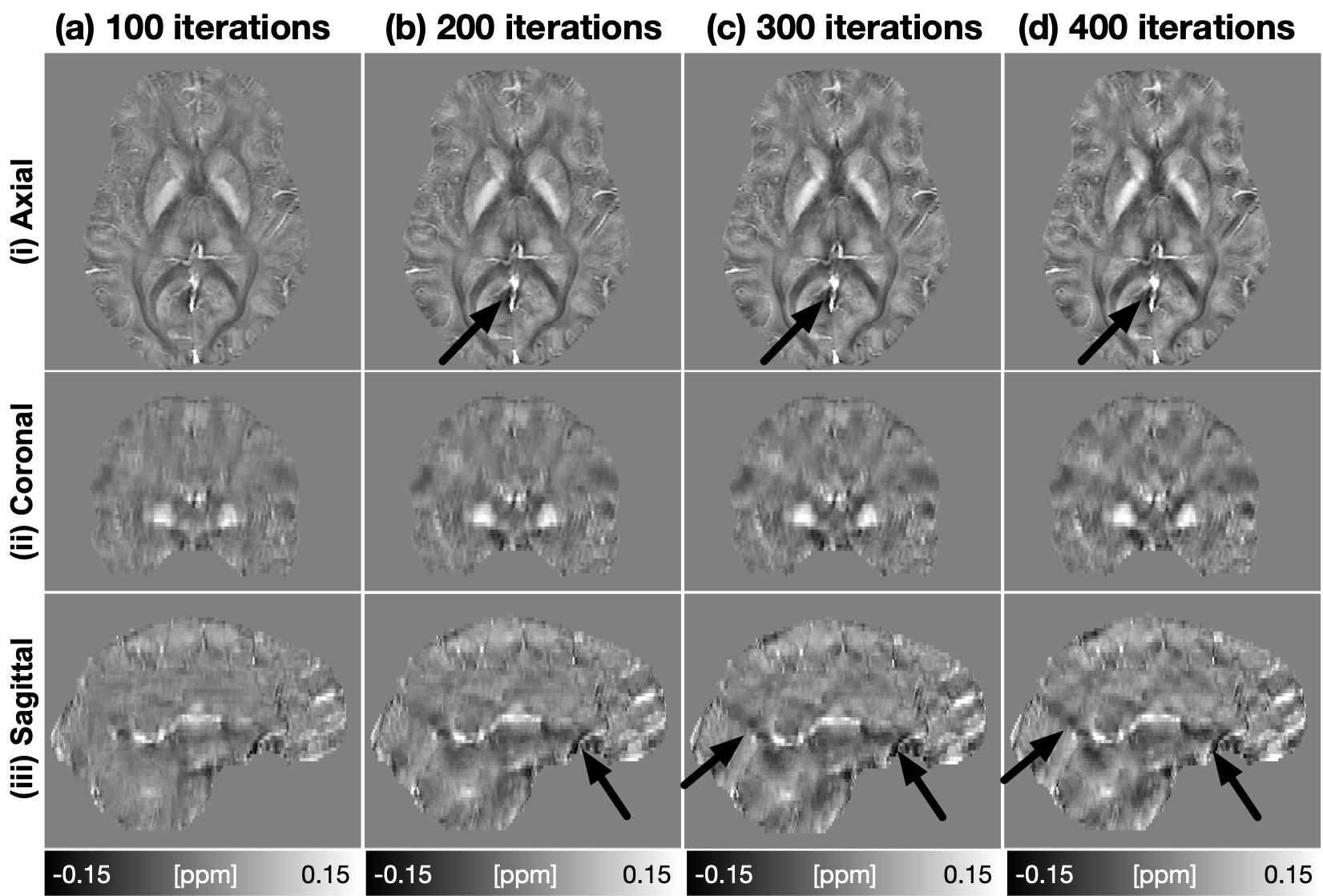}
\caption{Comparison of DIP results from a single-orientation dataset reconstructed after 100, 200, 300, and 400 iterations.}
\label{fig:uQSM_DIP_H2H_Steps_1}
\vspace{-20pt}
\end{center}
\end{figure}

\begin{figure}[H]
\vspace{-10pt}
\begin{center}
\includegraphics[width=\textwidth]{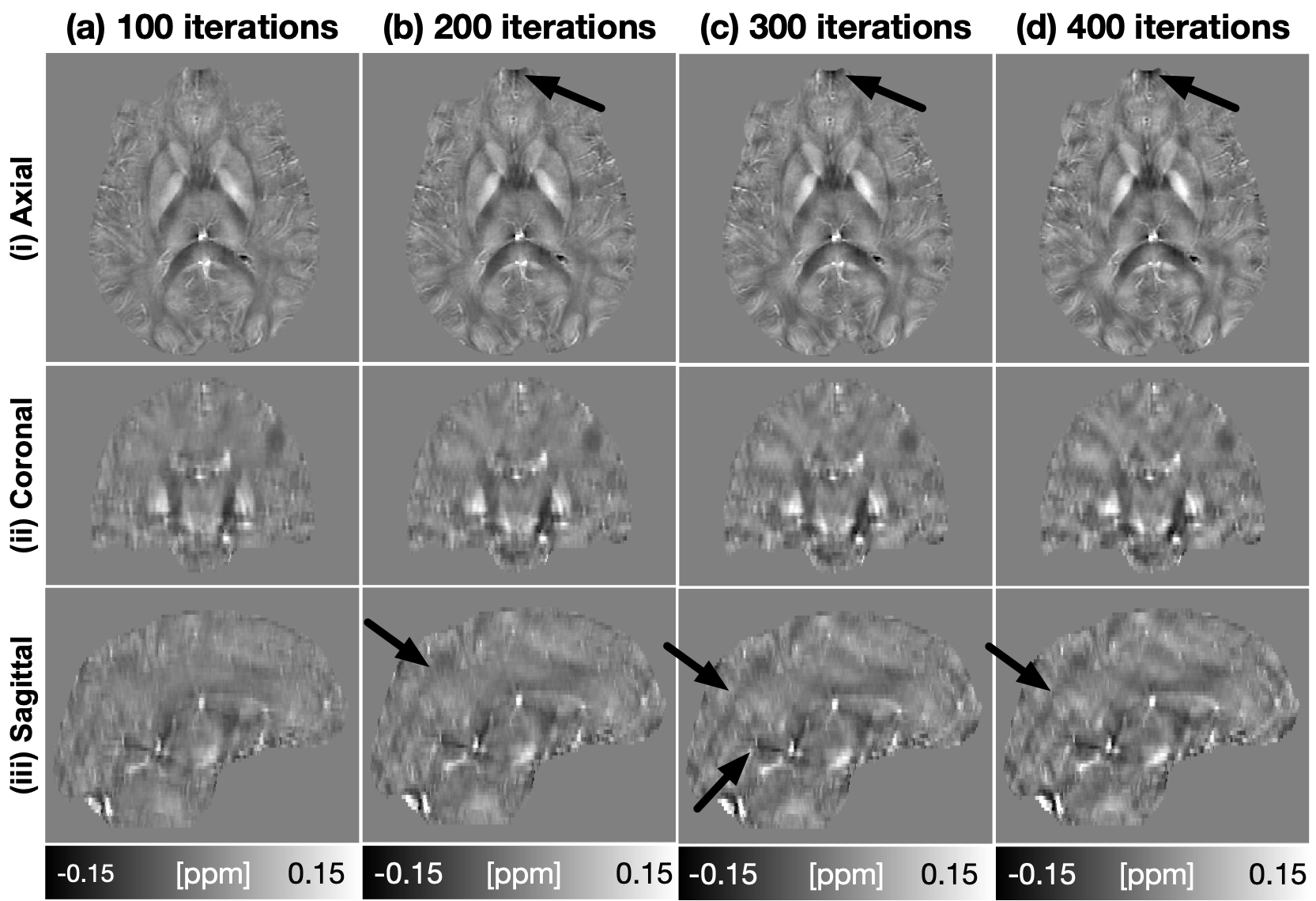}
\caption{Comparison of DIP results from a single-orientation dataset reconstructed after 100, 200, 300, and 400 iterations.}
\label{fig:uQSM_DIP_H2H_Steps_2}
\end{center}
\vspace{-20pt}
\end{figure}

Fig.\ref{fig:uQSM_DIP_H2H} displayed QSM images from two single-orientation datasets reconstructed by DIP and uQSM. DIP results showed inferior susceptibility maps with streaking artifacts while uQSM generated better susceptibility maps.

In order to better understand the overfitting problem in DIP, here we displayed the DIP results of 100, 200, 300, and 400 iterations in multi-orientation datasets and single-orientation datasets.

Fig.\ref{fig:uQSM_DIP_QSMdata_Steps_1} displayed the DIP results of 100, 200, 300, and 400 iterations on a multi-orientation dataset. It was clearly shown than DIP images of 100 iterations demonstrated inferior image contrast but invisible image artifacts. While DIP images of 200, 300, and 400 iterations exhibited more shading artifacts. 

Fig.\ref{fig:uQSM_DIP_H2H_Steps_1} and Fig.\ref{fig:uQSM_DIP_H2H_Steps_2} displayed the DIP results of 100, 200, 300, and 400 iterations from two multi-orientation dataset. It was clearly observed than DIP results of 100 iterations displayed inferior image contrast but less image artifacts. DIP results of 200 iterations results displaying high quality QSM images with subtle artifacts, while DIP results of 300 and 400 iterations results showed severe shading streaking artifacts due to overfitting. 

\section{Deconvolution and Checkboard Artifacts}
In uQSM network architecture, upsampling layers were used in stead of transposed convolutional layers, i.e., deconvolution, in order to reduce the checkerboard artifacts in the QSM results. 

\begin{figure}[H]
\begin{center}
\includegraphics[width=\textwidth]{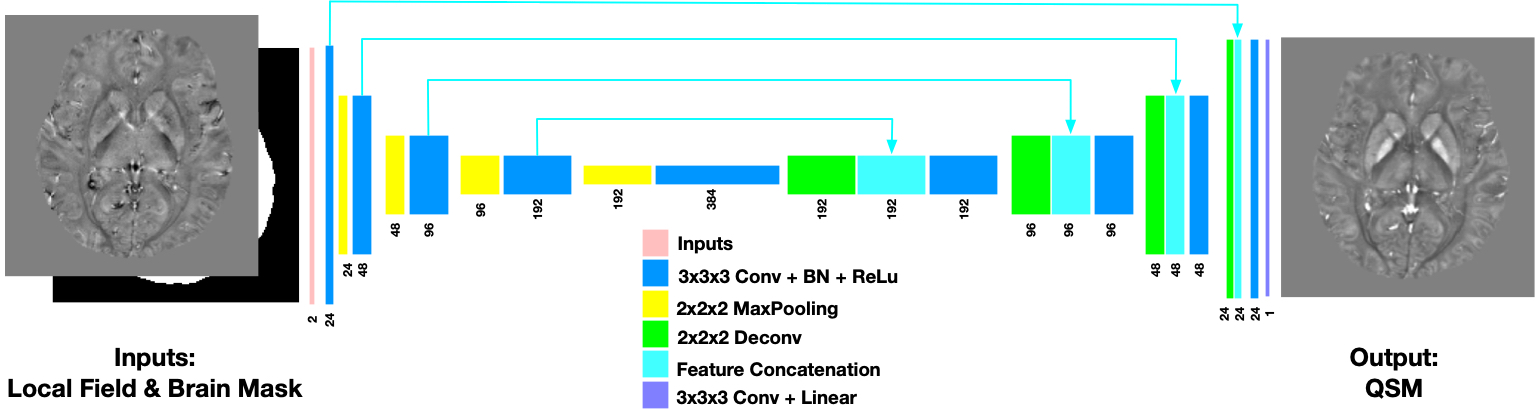}
\caption{Neural network architecture of uQSM using deconvolution. It has an encoder-decoder structure with 9 convolutional layers (kernel size 3x3x3, same padding), 9 batch normalization layers, 9 ReLU layers, 4 max pooling layers (pooling size 2x2x2, strides 2x2x2), 4 transposed convolutional layers (kernel size 2x2x2, size 2x2x2), 4 feature concatenations, and 1 convolutional layer (kernel size 3x3x3, linear activation).}
\label{fig:uQSM_CNN_Deconvolution}
\vspace{-20pt}
\end{center}
\end{figure}

We compared the network using upsampling layers with the network using deconvolutional layers. The neural network in Fig.\ref{fig:uQSM_CNN_Deconvolution} using deconvolution had 6,078,865 total parameters and 6,076,657 trainable parameters. The neural network using upsampling layers had 6,617,065 total parameters and 6,614,857 trainable parameters. These two neural networks had close number of parameters. 

\begin{figure}[H]
\begin{center}
\vspace{-15pt}
\includegraphics[width=\textwidth]{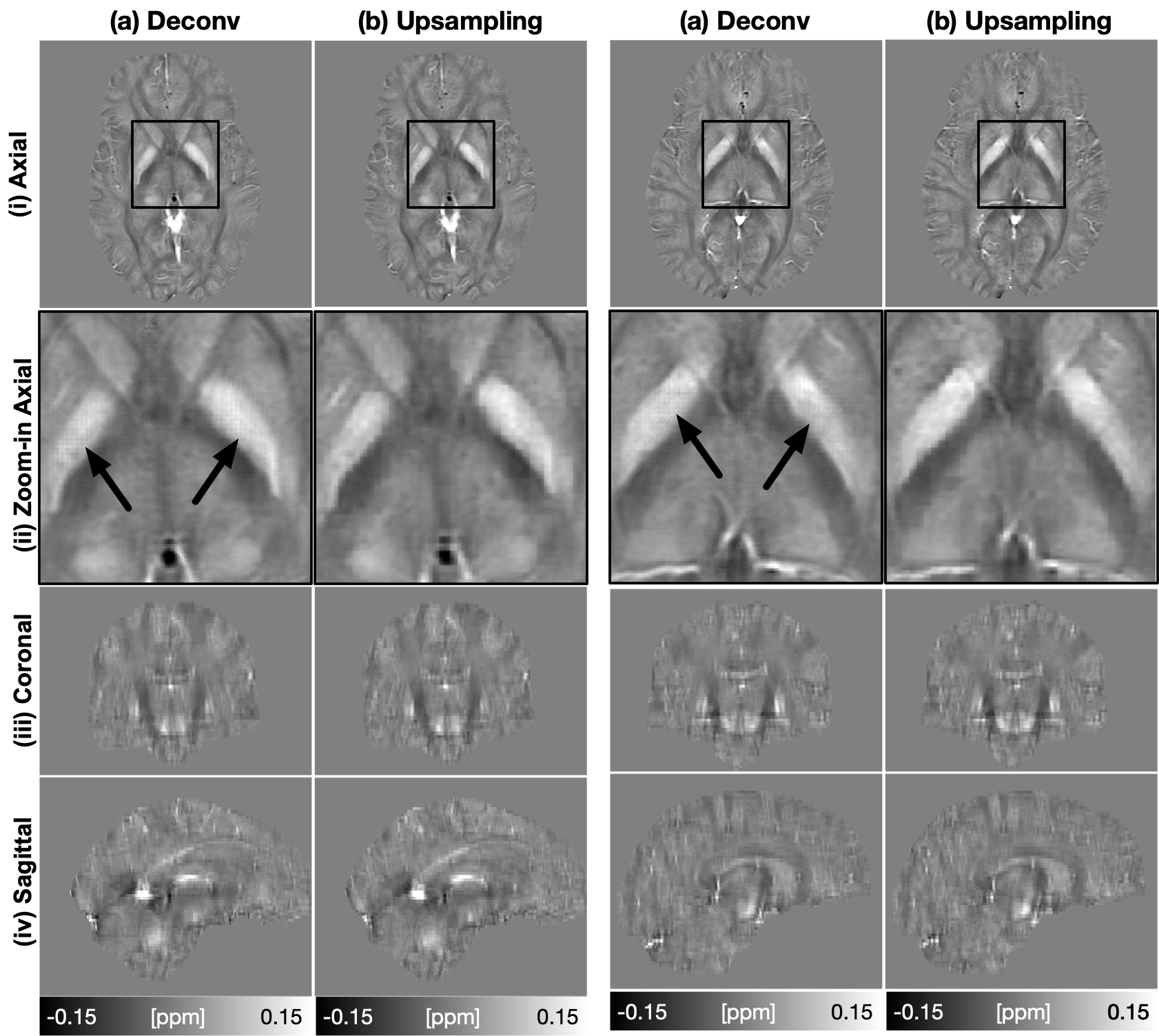}
\caption{Comparison of QSM results from 2 single-orientation datasets reconstructed with deconvolution-based network and upsampling-based network. The QSM of deconvolution-based network showed checkboard artifacts in the zoom-in axial plane (a, black arrows), while upsampling-based network suppressed these artifacts.}
\label{fig:uQSM_upSamplingH2H_1}
\vspace{-15pt}
\end{center}
\end{figure}

Fig.\ref{fig:uQSM_upSamplingH2H_1} displayed QSM images from 2 single-orientation datasets reconstructed deconvolution-based network and upsmapling-based network. It was observed that QSM of deconvolution-based network showed severe checkboard artifacts, while QSM of upsmapling-based network did not have checkboard artifacts.

\subsection{Effects of the Data Consistency Loss}
In the loss function of uQSM, the nonlinear dipole inversion data consistency loss was utilized in order to get more reliable QSM. Here, we compared three different data consistency losses. 

(1) Linear dipole inversion (LDI), 
\begin{equation}
L_{LDI} = \left \|m(d\ast \chi  -  y) \right \|_{2}
\end{equation}

(2) Weighted linear dipole inversion (WDI), 
\begin{equation}
L_{WLDI} = \left \|W m (e^{i d\ast \chi}  - e^{i y}) \right \|_{2}
\end{equation}

(3) Weighted nonlinear dipole inversion (NDI), 
\begin{equation}
L_{NDI} = \left \| W m (d\ast \chi  -  y) \right \|_{2}
\end{equation}

We trained uQSM with three different data consistency losses on 9 multi-orientation datasets. The training scheme was the same as aforementioned. Quantitative metrics, pSNR, NRMSE, HFEN, and SSIM, were calculated with COSMOS map as a reference.    

\begin{table}[H]
\centering
\caption{\label{tab:uQSM_Loss} Means and standard deviations of quantitative performance metrics from uQSM using 3 different data consistency losses on 9 multi-orientation datasets. NDI achieved the best scores in all criteria, WDI achieved the second best in all quantitative metrics, and LDI got the worst criterion performance in all metrics.}
\vspace{0in}
\begin{tabular}{cccccc}
\hline
\multicolumn{1}{|c}{} & \multicolumn{1}{|c}{pSNR (dB)} & \multicolumn{1}{|c}{NRMSE ($\%$)} &\multicolumn{1}{|c}{HFEN ($\%$)} &\multicolumn{1}{|c|}{SSIM (0-1)}\\
\hline 
\multicolumn{1}{|c}{LDI} & \multicolumn{1}{|c}{$44.1\pm0.5$} & \multicolumn{1}{|c}{$84.9\pm5.9$} & \multicolumn{1}{|c}{$73.4\pm5.7$} & \multicolumn{1}{|c|}{{$0.879\pm0.013$}}\\
\hline
\multicolumn{1}{|c}{WDI} & \multicolumn{1}{|c}{$45.0\pm0.5$} & \multicolumn{1}{|c}{$75.9\pm5.4$} & \multicolumn{1}{|c}{$67.1\pm5.5$} & \multicolumn{1}{|c|}{{$0.888\pm0.015$}}\\
\hline
\multicolumn{1}{|c}{NDI} & \multicolumn{1}{|c}{\textbf{45.6$\pm$0.4}} & \multicolumn{1}{|c}{\textbf{71.4$\pm$5.0}} & \multicolumn{1}{|c}{\textbf{62.8$\pm$5.0}} & \multicolumn{1}{|c|}{\textbf{0.890$\pm$0.015}}\\
\hline 
\vspace{-15pt}
\end{tabular}
\begin{flushleft}
\end{flushleft}
\end{table}

\begin{figure}[H]
\begin{center}
\vspace{-5pt}
\includegraphics[width=\textwidth]{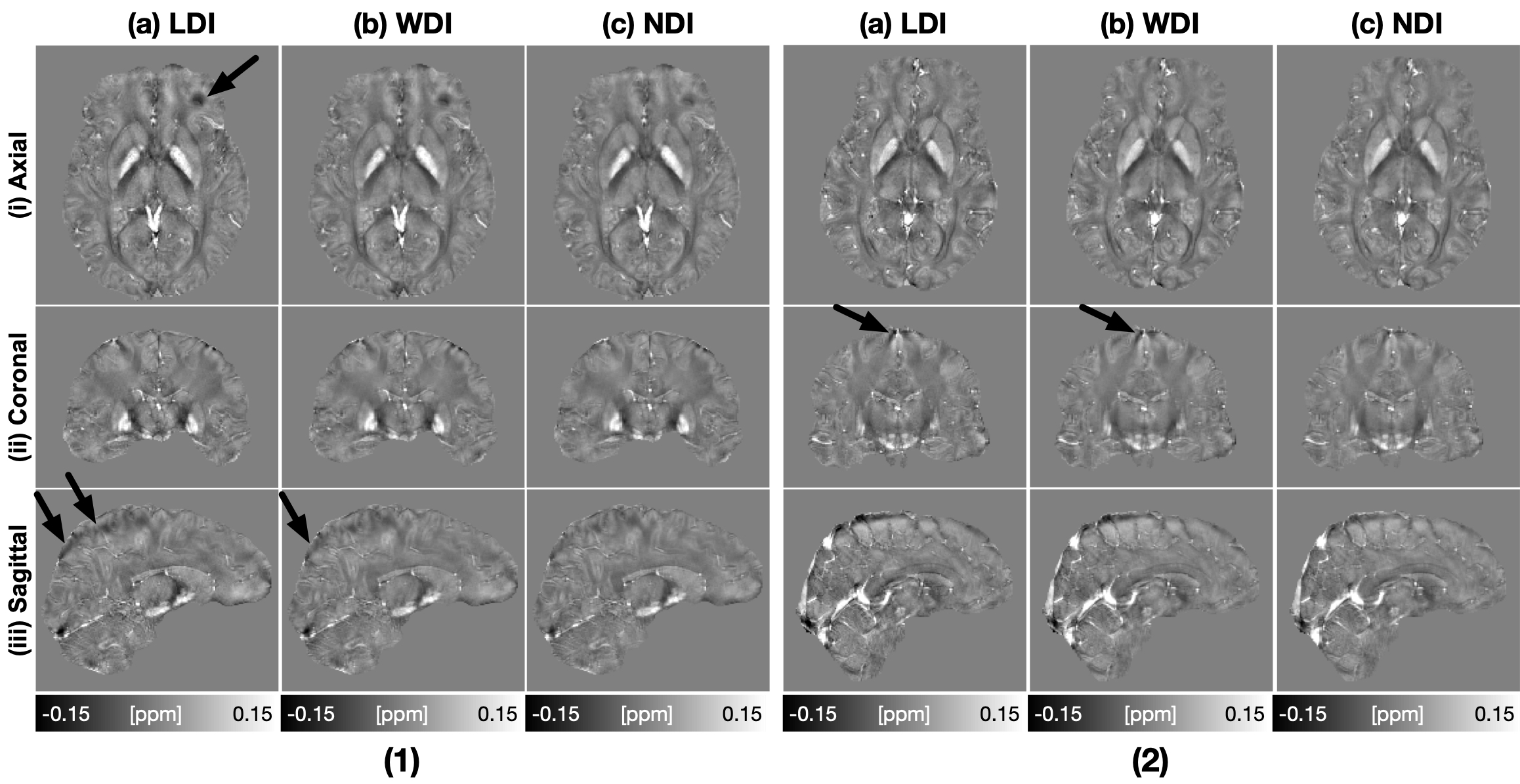}
\caption{Comparison of QSM results using different data consistency losses of 2 multi-orientation datasets. LDI (a) and WDI (b) images showed slight black shading artifacts (a-b, black arrows), while NDI (c) images better suppressed these shadings.}
\label{fig:uQSM_loss_QSMdata}
\vspace{-15pt}
\end{center}
\end{figure}

Table.\ref{tab:uQSM_Loss} displayed the quantitative metrics of uQSM results using 3 different data consistency losses on multi-orientation datasets. uQSM using $L_{NDI}$ achieved the best scores in pSRN, NRSE, HFEN, and SSIM, then $L_{WDI}$ achieved the second best in all quantitative metrics, and $L_{LDI}$ got the worst performance.    

Fig.\ref{fig:uQSM_loss_QSMdata} displayed the uQSM results of 2 multi-orientation datasets. LDI (a), WDI (b) images displayed black shading artifacts close to brain vessels with high susceptibility values (a-b, black arrows), while NDI (c) suppressed these artifacts and produced superior QSM images.

\begin{figure}[H]
\begin{center}
\vspace{-15pt}
\includegraphics[width=\textwidth]{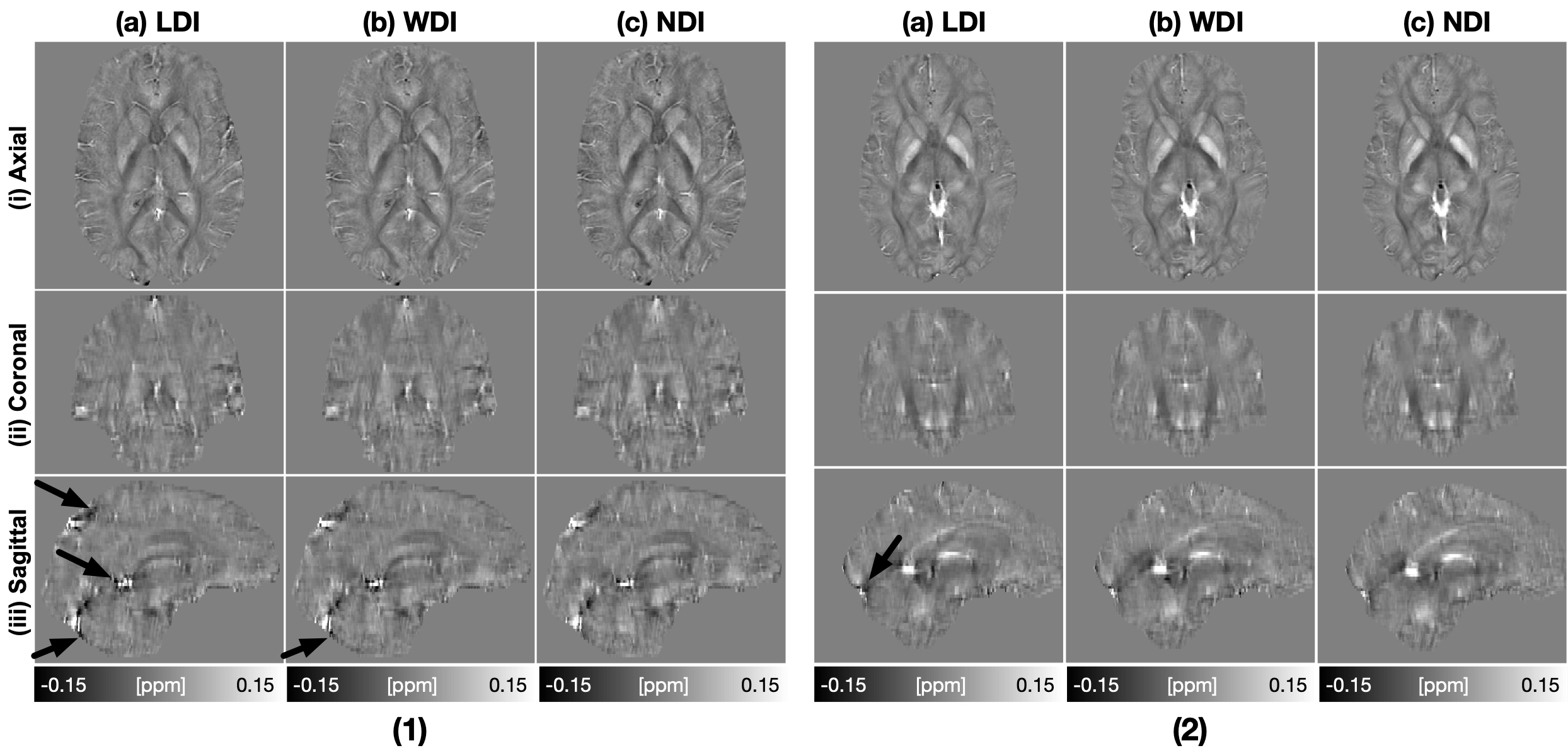}
\caption{Comparison of QSM results reconstructed using different data consistency losses of 2 single-orientation data. LDI (a), WDI (b) results showed black shading artifacts close to vessels (a, b, black arrows), while NDI (c) generated better QSM images with less artifacts.}
\label{fig:uQSM_loss_H2H}
\vspace{-15pt}
\end{center}
\end{figure}

Fig.\ref{fig:uQSM_loss_H2H} displayed the QSM results reconstructed using three different data consistency losses of 2 single-orientation data. It was observed that LDI (a), WDI (b) images displayed black shading artifacts close to brain vessels with high susceptibility values (a, b, black arrows), while NDI (c) suppressed these artifacts in the reconstructed QSM images.

When comparing three different data consistency losses, the linear dipole inversion loss and weighted linear dipole inversion loss are prone to introduce artifacts in the reconstructed QSM images in both multi-orientation datasets and single-orientation datasets, while nonlinear dipole inversion loss produces better QSM images. These results showed that nonlinear dipole inversion can achieve more robust QSM images since the local fields have unknown spatially variant noise. 

\subsection{Effects of TV $L_{TV}$}

In uQSM, $L_{TV}$ was incorporated in the loss function in order to suppress the noise and preserve edges in the reconstructed QSM estimates. Here, we investigated the effects of $L_{TV}$ in multi-orientation datasets which showed high level of noise due to that the dataset were acquired with high image resolution and short scan time per orientation. 

\begin{table}[H]
\centering
\caption{\label{tab:uQSM_tv} Means and standard deviations of quantitative performance metrics of uQSM with $L_{TV}$ and without $L_{TV}$ with the COSMOS map as a reference on 9 multi-orientation datasets. The quantitative evaluation showed that uQSM with $L_{TV}$ achieved better scores in all metrics.}
\vspace{0in}
\begin{tabular}{cccccc}
\hline
\multicolumn{1}{|c}{} & \multicolumn{1}{|c}{pSNR (dB)} & \multicolumn{1}{|c}{NRMSE ($\%$)} &\multicolumn{1}{|c}{HFEN ($\%$)} &\multicolumn{1}{|c|}{SSIM (0-1)}\\
\hline 
\multicolumn{1}{|c}{w/o $L_{TV}$} & \multicolumn{1}{|c}{$43.8\pm0.5$} & \multicolumn{1}{|c}{$87.4\pm6.8$} & \multicolumn{1}{|c}{$70.5\pm5.7$} & \multicolumn{1}{|c|}{{$0.848\pm0.022$}}\\
\hline
\multicolumn{1}{|c}{w $L_{TV}$} & \multicolumn{1}{|c}{\textbf{45.6$\pm$0.4}} & \multicolumn{1}{|c}{\textbf{71.4$\pm$5.0}} & \multicolumn{1}{|c}{\textbf{62.8$\pm$5.0}} & \multicolumn{1}{|c|}{\textbf{0.890$\pm$0.015}}\\
\hline 
\end{tabular}
\vspace{-15pt}
\begin{flushleft}
\end{flushleft}
\end{table}

Table.\ref{tab:uQSM_tv} displayed the quantitative metrics of uQSM results with $L_{TV}$ and with $L_{TV}$ on multi-orientation datasets. uQSM with $L_{TV}$ achieved the best scores in pSRN, NRSE, HFEN, and SSIM.

\begin{figure}[H]
\vspace{-15pt}
\begin{center}
\includegraphics[width=\textwidth]{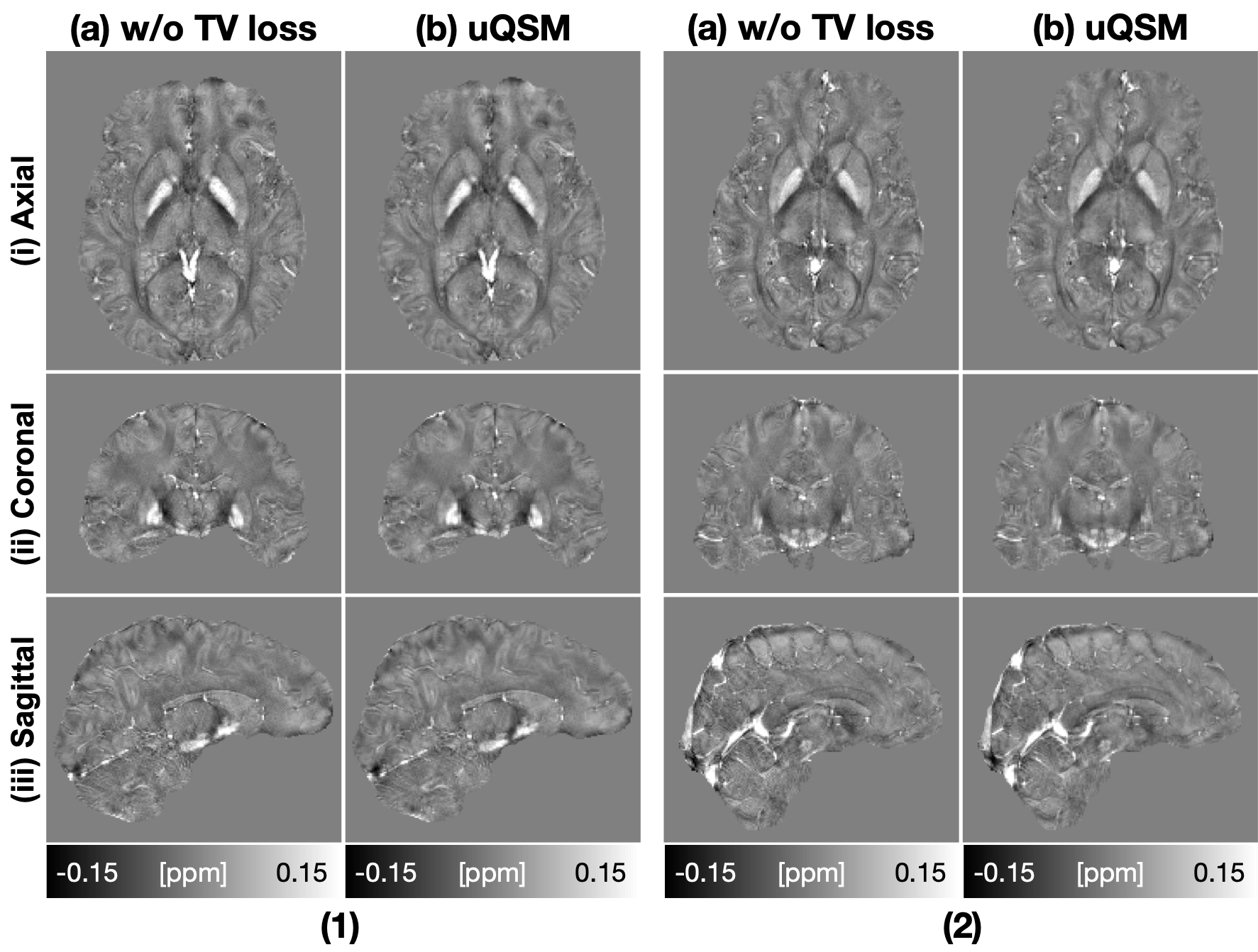}
\caption{Comparison of QSM results of uQSM with/without $L_{TV}$ on 2 multi-orientation datasets. uQSM without $L_{TV}$ showed high level of noise in the QSM results, while uQSM with $L_{TV}$ greatly reduced the noise and at same time preserved the fine details in the reconstructed QSM.}
\label{fig:uQSM_TV_QSMnet}
\vspace{-15pt}
\end{center}
\end{figure}

Fig.\ref{fig:uQSM_TV_QSMnet} displayed QSM results reconstructed using uQSM with/without $L_{TV}$ in neural network training of 2 multi-orientation datasets. QSM of uQSM without $L_{TV}$ showed high level of image noise, while uQSM with $L_{TV}$ better suppressed the noise and did not sacrifice image details in the reconstructed susceptibility maps.

When investigating the effects of TV loss, we found that uQSM trained with TV loss can greatly suppress the noise and preserve image details in the multi-orientation datasets. Since the multi-orientation datasets were acquired using a single echo GRE pulse sequence with a scan time of 1 minute at each orientation and high image resolution 1x1x1 mm$^3$, the datasets had high level of noise and low SNR. On the contrary, the single-orientation data were acquired using a four-echo GRE pulse sequence with a scan time of 4 minute and low image resolution. Thus, the single-orientation datasets have high SNR and it is necessary to incorporate the  $L_{TV}$ in the loss function for the single-orientation datasets.